\documentclass[twocolumn,showpacs,preprintnumbers,amsmath,amssymb,prb]{revtex4}

\usepackage{graphicx}% Include figure files
\usepackage{dcolumn}% Align table columns on decimal point
\usepackage{epsf}% bold math

\begin{document}

\title{Dynamics of vortex penetration, jumpwise instabilities and nonlinear surface resistance
of type-II superconductors in strong rf fields}

\author{A. Gurevich$^{1}$ and G. Ciovati$^{2}$}

\affiliation{$^{1}$National High Magnetic Field Laboratory, Florida
State University, Tallahassee, Florida 32310, \\
$^{2}$Thomas Jefferson National Accelerator Facility, Newport News, Virginia 23606}

\date{\today}

\begin{abstract}

We consider nonlinear dynamics of a single vortex in a superconductor in a strong rf
magnetic field $B_0\sin\omega t$. Using the London theory, we
calculate the dissipated power $Q(B_0,\omega)$, and the transient time scales of
vortex motion for the linear Bardeen-Stephen viscous drag force, which
results in unphysically high vortex velocities during vortex
penetration through the oscillating surface barrier. It is shown that
penetration of a single vortex through the ac surface barrier always involves
penetration of an antivortex and the subsequent annihilation of the vortex
antivortex pairs. Using the nonlinear Larkin-Ovchinnikov (LO) viscous drag force at higher
vortex velocities $v(t)$ results in a jump-wise vortex penetration through the
surface barrier and a significant increase of the dissipated power. We
calculate the effect of dissipation on nonlinear vortex viscosity $\eta(v)$ and
the rf vortex dynamics and show that it can also result in the LO-type behavior,
instabilities, and thermal localization of penetrating vortex channels. We propose a thermal feedback
model of $\eta(v)$, which not only results in the LO dependence of $\eta(v)$ for a steady-state motion, but
also takes into account retardation of temperature field around rapidly accelerating
vortex, and a long-range interaction with the surface.  We also address
the effect of pinning on the nonlinear rf vortex dynamics and the effect of trapped magnetic flux on
the surface resistance $R_s$ calculated as a function or  rf frequency and field. It is shown that
trapped flux can result in a temperature-independent residual resistance $R_i$ at low $T$, and a hysteretic low-field
dependence of $R_i(B_0)$, which can {\it decrease} as $B_0$ is increased, reaching a minimum at
$B_0$ much smaller than the thermodynamic critical field $B_c$. We propose that cycling of rf field can
reduce $R_i$ due to rf annealing of magnetic flux which is pumped out by rf field from the
thin surface layer of the order of the London penetration depth.

\end{abstract}

\maketitle

\section{Introduction}

The behavior of superconductors in strong rf fields involves many
complex mechanisms related to a nonlinear electromagnetic response
of nonequilibrium quasiparticles, pairbreaking
suppression of superconducting gap $\Delta$ and penetration of
vortices at higher rf amplitudes  \cite{bardeen,kopnin}. The physics behind the
nonlinear rf response has recently attracted much attention due
to the development of a new generation of high-performance superconducting Nb
cavities for particle accelerators, in which the peak surface GHz fields
$B(t)=B_0\sin\omega t$ close to the thermodynamic critical field $B_c$ were reached at a
very high quality factor $\sim 10^9-10^{11}$ characteristic of the
Meissner state \cite{limit,hasan}. At such strong rf fields the peak
surface current density $B_0/\mu_0\lambda$ approaches the depairing
current density $J_d$ at which the Meissner state becomes
unstable with respect to avalanche penetration of vortices once the
instantaneous rf field $B(t)=B_0\sin\omega t$ exceeds the
superheating field $B_s\approx B_c$. In turn, penetration of
vortices causes a sharp increase in the surface resistance $R_s$.

As far as the very high quality factors are concerned, of particular interest is the behavior of $R_s$ in
s-wave superconductors at low temperatures
$T\ll T_c$ and frequencies $\omega\ll\Delta$, for which the rf field cannot break
the Cooper pairs, and the very low Meissner surface resistance $R_s\propto
(\omega^2\Delta/T)\exp(-\Delta/T)$ is due to an exponentially small density of
thermally-activated quasiparticles (unlike the power-law dependence $R_s(T)\simeq R_i+CT^\alpha$
due to nodal quasiparticles in d-wave superconductors \cite{rshts1,rshts2,rshts3,rshts4}).  In this case penetration
of even few vortices driven by extremely high rf currents densities $J\sim J_d$ can produce strong energy dissipation
comparable to that in the Meissner state, which in turn, can trigger thermomagnetic flux avalanches and the
superconductivity breakdown. It is therefore important to understand mechanisms, which control dynamics of single
vortex penetration under strong rf fields.   Yet, the rf field onset of vortex penetration $B_v$, and the
dissipated power $Q$ as functions of $B_0$ and $\omega$, and the relation between $B_v$ and the thermodynamic $B_c$ and the
lower critical field $B_{c1}$ are still not well understood. These problems
include complex kinetics of the emergence of the vortex core at the surface, and the
subsequent nonlinear large-amplitude oscillation of the vortex at
the surface driven by strong rf currents much higher than depinning
critical current density. This situation cannot be described by a well
developed linear electrodynamics of a superconductor in the pinned mixed
state weakly deformed by rf currents \cite{imped1,imped2,imped3,imped4}.
Some issues of nonlinear vortex dynamics in ramping magnetic fields
have been addressed in extensive numerical simulations of the
time-dependent Ginzburg-Landau (TDGL) equations \cite{tdgl1,tdgl2,tdgl3,tdgl4} valid at $T\approx T_c$,
and molecular dynamics simulations \cite{md1}. However, few
experimental and theoretical results on vortices driven by very
strong rf currents at low temperatures have been published in the
literature.

In this paper we address nonlinear rf dynamics of a single vortex
moving in and out of a type-II superconductor through an oscillating
magnetic surface barrier locally weakened by a surface defect. We
show that already this basic problem involves a rich physics, since
even weak Meissner fields $B_0\ll B_c$ can drive the vortex with
velocities $v(t)$ so high that the linear Bardeen-Stephen viscous
drag model becomes inadequate. As a result, the vortex velocity
$v(t)$ can exceed the sound velocity, causing the Cherenkov
generation of hypersound \cite{vorts1,vorts2}. Moreover, $v(t)$ can
exceed the critical velocity $v_0$, above which the vortex drag
coefficient $\eta(v)$ decreases as $v$ increases, and the viscous
drag force $f_v = v\eta(v)$ reaches maximum at the critical velocity
$v_0$, resulting in the jumpwise Larkin-Ovchinnikov (LO) instability
\cite{lo,bs}. The LO instability has been extensively investigated
by dc transport measurements
\cite{inst1,inst2,inst3,inst4,inst5,inst6,inst7} on both low-T$_c$
and high-T$_c$ superconductors for which $v_0\simeq 1-10$ km/s have
been typically observed at low $T$ and $B$. Single-vortex dynamics
under strong rf field also involves annihilation of
vortex-antivortex pairs, and a cascade of single, double and
multiple vortex penetrations. Competition of the rf driving force,
image attraction to the surface, and the viscous drag force results
in a strong dependence of the dissipated power $Q$ on the rf
amplitude and frequency. Very high vortex velocities achieved at
fields $B_0\sim B_c$ required to break the surface barrier make it
possible to probe the behavior of vortices under extreme conditions,
for which the Lorentz driving force approaches its ultimate
depairing limit. Because of strong heating effects, these conditions
are hard to reproduce in dc transport experiments (except in
high-power pulse measurements \cite{pulse1,pulse2}). The importance
of heating effects for transport instabilities in superconductors at
low temperatures is well known \cite{bs,gm, heat1,heat2}. In this
paper we show that heating is a key limiting factor for the
high-field surface resistance at $T\ll T_c$ as well, even for single
vortices driven by strong rf Meissner currents. In particular,
viscous vortex dynamics coupled with electron overheating can result
in the LO-type behavior of $\eta(v)$, thermal rf breakdown,
long-range interaction (on scales much greater than the London
penetration depth) between a vortex and the surface and between
vortices themselves.

The paper is organized as follows. In section II we establish the
main parameters of interest by considering
penetration and dissipation of a single vortex over the oscillating
surface barrier in type-II superconductors described by the dynamic equation, in which the linear
Bardeen-Stephen viscous drag force is balanced by the Lorentz driving
force and the image attraction force at the surface in the London theory. Even in
this basic model rf vortex dynamics always involves annihilation
of vortex-antivortex pairs for  $B_0\approx B_v$ close to the penetration field $B_v$ and a strong dependence of the
dissipated power $Q(B_0,\omega)$ on the rf frequency and amplitude. In section-III we
show that the Bardeen-Stephen model actually has a very limited
applicability because vortices breaking through the surface barrier reach supersonic velocity so
the dependence of the viscous drag coefficient
$\eta(v)$ must be taken into account. In this case vortex dynamics
becomes strongly coupled with nonequilibrium overheating of the
vortex core, resulting in jumpwise penetration of single vortices
through the surface barrier, and significant increase in $Q$. In
section IV we consider the effect of pinning on rf
surface resistance. In particular, we show that trapped vortices can result in
a temperature independent, field-hysteretic residual resistance, which
can {\it decrease} as the rf field increases. Pinned vortices can also produce
hotspots, which ignite thermal rf breakdown.  Section V is devoted to dissipation around
hotspots and their nonlinear contribution to the global surface resistance. The thermal breakdown
of the Meissner state ignited  by vortex hotspots is addressed.
Section VI concludes with a discussion of the results.

\section{ Penetration of a vortex over the oscillating surface barrier}

\subsection{Dynamic equations and time scales}

    \begin{figure}                  %FIG2
    \epsfxsize= 0.43\hsize
    \centerline{
    \vbox{
    \epsffile{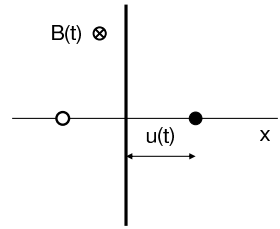}
    }}
    \caption{Vortex (shown as the filled circle) penetrating by the distance $u(t)$ from the semi-infinite
    surface $(x=0)$ exposed to a uniform parallel rf field $B(t)$. The open circle
    shows the position of the antivortex image.
    }
    \label{Fig.1}
    \end{figure}

Penetration of vortices in a superconductor is controlled by
the Bean-Livingston surface barrier, which results from a
competition between the Meissner screening currents pushing the
vortex in a superconductor and the attraction force between a
vortex and the surface \cite{bl}. This surface barrier oscillates
under the rf field, so motion of a vortex in and out of a superconductor is described
by a dynamic equation. We consider here a type-II
superconductor in the London theory, assuming that the rf
field $B(t)=B_0\sin\omega t$ of amplitude $B_0$ and frequency
$\omega$ is applied parallel to the flat surface at $x>0$ as shown in Fig. 1. Then the
equation of motion for a single vortex driven by the rf Meissner
balanced by the image attraction force and the viscous drag force takes the form
    \begin{equation}
    \eta_0 \dot{u}=\frac{\phi_0B_0}{\mu_0\lambda}e^{-u/\lambda}\sin\omega t-\frac{\phi_0^2}{2\pi\mu_0\lambda^3}
    K_1\!\left[\frac{2}{\lambda}\sqrt{u^2+\xi_s^2}\right]\!,
    \label{dyn1}
    \end{equation}
where $u(t)$ is the distance of the vortex core from the surface,
$\lambda$ is the London penetration depth, $\eta_0 =
\phi_0B_{c2}/\rho_n$ is the Bardeen-Stephen vortex viscosity,
$\rho_n$ is the normal state resistivity, $\phi_0$ is the magnetic
flux quantum, $B_{c2} = \phi_0/2\pi\xi^2$ is the upper critical
field, $K_1(x)$ is the modified Bessel function. Here we introduce
the local coherence length $\xi_s$ at the point of the vortex entry,
which provides the cutoff in the London theory. For $u<\xi_s$, the
last term in Eq. (\ref{dyn1}) gives a constant force of vortex
attraction to the surface due to the formation of a "core string" of
depressed order parameter revealed by computer simulations of the GL
equations \cite{tdgl4,sh4}.

In this work we treat the emergence of the vortex
phenomenologically, assuming that it first appears in a small defect
region at the surface. For the results presented below, the actual
nature of the defect is not important as long as the defect size is
much smaller than $\lambda$, and the local $\xi_s$ is larger than
the bulk coherence length $\xi$. The vortex penetrates at the field
$B(t)>B_v$ for which the local surface barrier disappears because
the peak Meissner force $\phi_0B_0/\mu_0\lambda$ exceeds the maximum
attraction force to the surface
$\phi_0^2K_1(2\xi_s/\lambda)/2\mu_0\lambda^3$. For
$\xi_s\ll\lambda$, we can expand $K_1(x)\approx 1/x$, and obtain
    \begin{equation}
    B_v =\phi_ 0/4\pi\lambda\xi_s\approx 0.71B_c,
    \label{bv}
    \end{equation}
which basically defines $\xi_s$ in terms of the observed local
penetration field $B_v$, which has been calculated for different
types of surface defects \cite{sd1,sd2,sd3,sd4}.  We assume that
there is a distribution of sparse small regions with reduced local
$B_v$ on the surface where vortices first enter. Penetration of
straight vortices can only be initiated by linear defects (for
example, dislocations or grain boundaries) parallel to the vortex
line. For more common 3D surface defects, such as precipitates or
local variation of chemical composition, a vortex first emerges as a
semi-loop, which then expands as illustrated by Fig. 2. The initial
penetration of a curved vortex in Fig. 2 can hardly be described by
Eq. (\ref{dyn1}) for a straight vortex parallel to the surface.
However, the circular vortex semi-loop very quickly evolves into a
loop strongly elongated along the surface because of the gradient of
the Meissner current $J(x)=(B_0/\mu_0\lambda)\exp(-u/\lambda)$ and
the LO instability, which effectively straightens the vortex due to
jump-wise lateral propagation of the loop, as shown below. Thus, Eq.
(\ref{dyn1}) can be used after a short transient time, which is
still much smaller than the rf period.

     \begin{figure}                  %FIG2
    \epsfxsize= 0.45\hsize
    \centerline{
    \vbox{
    \epsffile{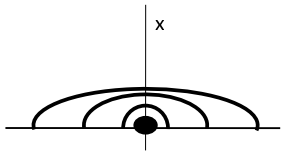}
    }}
    \caption{Snapshots of expanding vortex semi-loop emerging from a surface defect (black dot). The quicker expansion of the loop
    along the surface is due to the gradient of the Meissner current $J(x)\propto \exp(-u/\lambda)$ and the LO instability.
    }
    \label{Fig.2}
    \end{figure}

Eq. (\ref{bv}) gives $B_v$ close to the superheating field
$B_s=0.745 B_c$, above which the Meissner state in extreme type-II
superconductors with $\kappa=\lambda/\xi\gg 1$ becomes absolutely
unstable with respect to weak periodic perturbations of the order
parameter \cite{sh1,sh2,sh3,sh4,sh5,sh6} as the Meissner current
density at the surface $B_v/\mu_0\lambda$ exceeds the local
depairing current density $J_d$. For $B_0> B_v$, a vortex moves in
and out the superconductor under the action of the rf field. Since
Eq. (\ref{dyn1}) can only be used on the scales $u(t)>\xi_s$, we
neglect here a possible dependence of $\eta_0$ on $u$, although this
dependence can occur if a long-range interaction of the vortex core
with the surface due to nonequilibrium effects is taken into
account, as shown below.

To estimate the scale of vortex oscillations and maximum velocities,
we first disregard the image attraction force, which becomes
negligible at distances $u>\lambda$. Then the solution of Eq.
(\ref{dyn1}) takes the form
    \begin{equation}
    u(t)=\lambda\ln\left[1+\frac{\phi_0B_0}{\lambda^2\omega\eta_0\mu_0}(\cos\omega t_0-\cos\omega t)\right],
    \label{ex1}
    \end{equation}
where $t_0=\sin^{-1}(B_v/B_0)/\omega$ is the time of vortex entry. The maximum vortex
penetration depth $u_m$  corresponds to $\cos\omega t = -1$, whence
    \begin{equation}
    u_m=\lambda\ln\left[1+\frac{\phi_0}{\lambda^2\omega\eta_0\mu_0}\bigl(\sqrt{B_0^2-B_v^2}+B_0\bigr)\right]
    \label{amp}
    \end{equation}
Here $u_m$ depends logarithmically on the rf field and
frequency. From Eq. (\ref{ex1}) we estimate the time $\tau$ for
the vortex to move by the distance $\simeq \lambda$ from the
surface. For GHz frequencies and the materials parameters of Nb and Nb$_3$Sn,
$\tau$ turns out to be much shorter than the rf period so
$\cos\omega (t_0 + \tau)\simeq  \cos\omega t_0 -
\omega\tau\sin\omega t_0$, hence,
    \begin{equation}
    \tau=\frac{\mu_0\lambda^2\eta}{\phi_0B_v}\simeq\frac{2\mu_0\lambda^3}{\rho_n\xi}.
    \label{tau}
    \end{equation}
Taking $\rho_n = 0.2\mu\Omega$m,  $\lambda = 90$ nm, $\xi=3$ nm for
Nb$_3$Sn \cite{nb3sn}, we obtain $\tau\simeq 3\times 10^{-12}$ s and
$\omega\tau \simeq 0.04$ for 2GHz. Likewise, taking $\rho_n=10^{-9}
\Omega$m and $\lambda=\xi=40$ nm for Nb, yields $\tau=4\times
10^{-12}$s. Eq. (\ref{tau}) gives the lower limit for $\tau$ because
the image vortex attraction increases $\tau$.

Now let us consider rf vortex dynamics in more detail. During the
positive rf half-period, the vortex penetration starts once $B(t)$
exceeds the local $B_v$. Because the vortex currents flow
antiparallel to the Meissner currents at the surface, penetration of
the vortex suppresses the local pairbreaking instability. At the
time when the rf field almost changes sign, the vortex penetrates by
the maximum distance $u_m$ then it turns around and starts coming
back. However, for negative $B(t)$, the current density at the
surface $J(0,t)$ is now a sum of the vortex currents and the
parallel Meissner rf currents. As a result, when the outgoing vortex
reaches the critical distance $u_c$ from the surface, $J(0,t)$
exceeds $J_d$, causing penetration of an antivortex before the
vortex exits. The antivortex is driven into the sample by the
Meissner current and by the attraction to the outgoing vortex, to
which it annihilates at the distance $u_a$ from the surface. After
that the negative $B(t)$ reaches - $B_v$, and a new antivortex
penetrates the sample in the same way as the vortex did for the
positive cycle, except that once the antivortex reaches $u_c$  on
the way out, it creates a vortex at the surface, both annihilating
at $u_a$. This process repeats periodically.

Eq. (\ref{dyn1}) therefore describes vortex penetration and exit
until the Meissner current density plus the current density of the
outgoing vortex being at $x=u_c$ reaches -$B_v/\mu_0\lambda$ at the
surface at the time $t = t_c$ defined by:
    \begin{equation}
    B_0|\sin\omega t_c|+\frac{\phi_0}{\pi\lambda^2}K_1\left(\frac{u_c}{\lambda}\right)=B_v
    \label{tc}
    \end{equation}
The second term in the l.h.s. of Eq. (\ref{tc}) is twice the current
density of the vortex at the distance $u_c$ in an infinite sample
because the outgoing vortex and its antivortex image contribute
equally to $J(0,t)$. For $t
> t_c$, the vortex with the coordinate $u_+(t)$ and the antivortex
with the coordinate $u_-(t)$ move toward each other, as described by
the following equations:
    \begin{eqnarray}
    \eta_0\dot{u_+}=\frac{\phi_0B_0}{\mu_0\lambda}e^{-\frac{u_+}{\lambda}}\sin\omega t
    -\frac{\phi_0^2}{2\pi\mu_0\lambda^3}\bigl[
    K_1\left(\frac{2u_+}{\lambda}\right) \qquad
    \nonumber \\
    +K_1\left(\frac{u_+-u_-}{\lambda}\right)-K_1\left(\frac{u_++u_-}{\lambda}\right)\bigr],\qquad\qquad
    \label{dyn2} \\
    \eta_0\dot{u_-}=-\frac{\phi_0B_0}{\mu_0\lambda}e^{-\frac{u_-}{\lambda}}\sin\omega t-
    \frac{\phi_0^2}{2\pi\mu_0\lambda^3}\bigl[K_1\!\!\left(\frac{2}{\lambda}\sqrt{u_-^2+\xi_s^2}\right)
    \nonumber \\
    -K_1\left(\frac{u_+-u_-}{\lambda}\right)-K_1\left(\frac{u_++u_-}{\lambda}\right)\bigr]\qquad\qquad
    \label{dyn3}
    \end{eqnarray}
These equations reflect the balance of interaction forces between
the vortex and antivortex and their corresponding images similar to those in
Fig. 1. The first term in the r.h.s. of Eq. (\ref{dyn3}) has the minus sign because the Meissner current
drives the antivortex in the opposite direction as compared to the
vortex. The initial conditions for Eqs. (\ref{dyn2}) and (\ref{dyn3}) are: $u_+(t_c) =
u_c$, $u_-(t_c) = 0$; and the condition $u_+(t_a) = u_-(t_a) = u_a$
defines the annihilation distance $u_a$ and time $t_a$.

   \begin{figure}                  %FIG2
    \epsfxsize= 0.65\hsize
    \centerline{
    \vbox{
    \epsffile{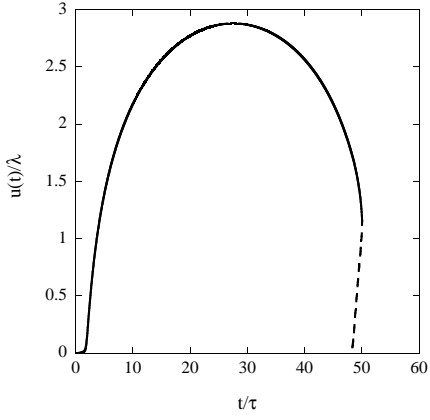}
    }}
    \caption{ Dynamics of vortex penetration and exit calculated from Eq. (\ref{dyn1}) for $t<t_c$
    and Eqs. (\ref{dyn2}) and (\ref{dyn3}) for $t>t_c$ for $\omega\tau =
    0.325$ and $B_0=1.02B_v$. The solid and dashed curves show the trajectories
    of the vortex and antivortex, respectively.
    }
    \label{Fig.3}
    \end{figure}

Dynamics of vortex penetration and annihilation is illustrated by
Fig. 3 where the vortex penetration depth is $u_m \simeq 3\lambda$,
the critical distance is $u_c\simeq2\lambda$, and the
vortex-antivortex annihilation occurs at $u_a\simeq \lambda$. For
$\omega\tau\ll 1$, the vortex first accelerates rapidly, penetrating
by the distance $\simeq \lambda$ during a time $\sim \tau$, and then
slowly turns around during the time of the order of the rf period
and annihilates in a short time $\sim\tau$.

     \begin{figure}                  %FIG3
    \epsfxsize= 0.65\hsize
    \centerline{
    \vbox{
    \epsffile{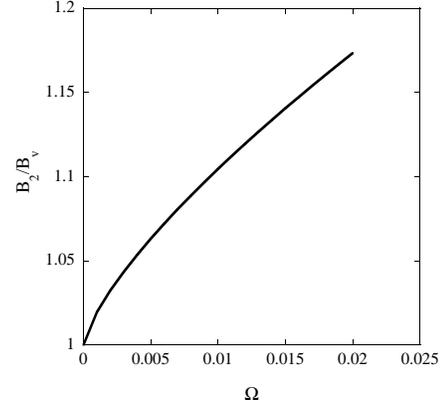}
    }}
    \caption{ Dependence of the second vortex penetration field $B_2$
    on the dimensionless frequency $\Omega=\omega\tau/\kappa$.
    }
    \label{Fig.4}
    \end{figure}

The above results are limited to the field region $B_v<B_0<B_s$
where single vortices penetrate independently through regions where
the Bean-Livingston barrier is locally suppressed by surface defects
separated by distances $> \lambda$. The case $B_0>B_s$ corresponds
to a global pairbreaking instability causing multi-vortex avalanche
penetration. Yet, even for $B_v<B_0<B_s$, a multi-vortex chain
penetration is possible. Indeed, penetration  of a single vortex for
$B(t)>B_v$ suppresses the local pairbreaking instability at $x=0$.
However, as $B(t)$ increases, the Meissner current density
increases, while the counterflow of surface current density at $x=0$
from the vortex decreases as it moves further away from the surface.
As a result, $J(0,t)$ can again reach $J_d$, causing penetration of
the second vortex at $t=t_2$ when the first vortex is located at
$x=u_2$ The condition of the second vortex penetration is similar to
Eq. (\ref{tc}),
    \begin{equation}
    B_2\sin\omega t_2-\frac{\phi_0}{\pi\lambda^2}K_1\left(\frac{u_2}{\lambda}\right)=B_v
    \label{bb2}
    \end{equation}
except for the minus sign in the l.h.s.  Eqs. (\ref{dyn1})
and (\ref{bb2}) define the critical rf amplitude $B_2$ below which
only the single vortex penetration occurs. Shown in Fig. 4 is the
curve $B_2(\omega)$ obtained by the numerical solution of Eqs.
(\ref{dyn1}) and (\ref{bb2}) for Nb$_3$Sn. These
results can be described well by the power law dependence
    \begin{equation}
    B_2=[1+ p\Omega^{\alpha}]B_v, \qquad \Omega=2\mu_0\lambda^2\omega/\rho_n,
    \label{b2a}
    \end{equation}
where $\Omega = \omega\tau/\kappa$, $\alpha=0.73$ and $p=0.23$. For
$\Omega\ll 1$, the field $B_2$ is close to $B_v$, however,
dissipation produced by penetrating vortices can significantly
reduce both $B_v$ and $B_2$ (see below).

\subsection{Vortex dissipation}

    \begin{figure}                  %FIG3
    \epsfxsize= 0.75\hsize
    \centerline{
    \vbox{
    \epsffile{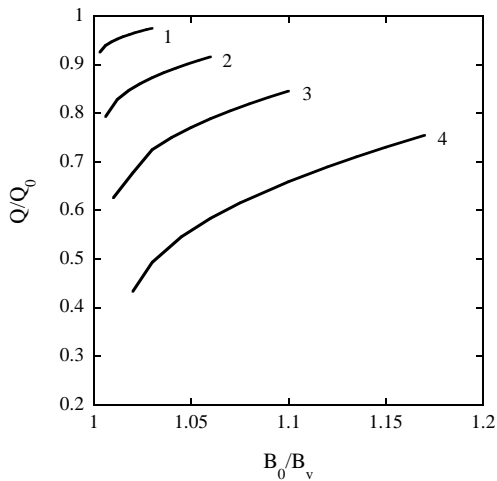}
    }}
    \caption{ Dissipated power Q as a function of the rf amplitude
    in the region of the single-vortex penetration $B_v<B_0<B_2$ for
    different frequencies $\omega\tau$: 0.061 (1); 0.162 (2); 0.325 (3); 0.65 (4).
    Here $Q_0=2\omega\phi_0B_v/\pi\mu_0$.}
    \label{Fig.5}
    \end{figure}
The power $Q=(\omega\eta/2\pi)\oint v^2 dt$ dissipated due to the work of the viscous drag forces
is given by
    \begin{equation}
    Q=\frac{\omega\eta}{\pi}\left[\int_{t_0}^{t_c}{\dot u}^2dt+\int_{t_c}^{t_a}({\dot u_+}^2+{\dot u_-}^2)dt\right],
    \label{w}
    \end{equation}
where $u(t)$ is the solution of Eq. (\ref{dyn1}), which describes
dynamics of a single vortex driven by rf field until $t=t_c$ when
the antivortex appears. The second integral in Eq. (\ref{w}) is due
to the collapse of the vortex-antivortex pair described by Eqs.
(\ref{dyn2}) and (\ref{dyn3}). For a quasistatic field, $Q$ can be obtained from the
change of the vortex thermodynamic potential $G(u)$
    \begin{equation}
    G(u)=\frac{\phi_0}{\mu_0}\left[B_{c1}-B+Be^{-u/\lambda}-\frac{\phi_0}{4\pi\lambda^2}K_0\!\left(\frac{2u}{\lambda}\right)\right]
    \label{gu}
    \end{equation}
where $B_{c1} = \epsilon\mu_0/\phi_0$ is the lower critical field
and $\epsilon$ is the vortex self-energy. If the ac field $B(t)$
varies very slowly $(\Omega\ll 1)$, the dissipated energy equals the
sum of $\Delta G_+=G(0)-G(u_m)$ during the positive half-cycle and
$\Delta G_-=G(u_m)-G(0)$ during the negative half-cycle. For any
closed vortex trajectory, which starts and ends at the surface,
contributions to $Q$ due to vortex self-energy and the work $\oint
F_i(u)\dot{u}dt$ of the potential image force $F_i(u)$ vanish. Thus,
$Q$ is only determined by the work of the driving Lorentz force,
$\simeq (2\phi_0B_v/\mu_0)[1-\exp(-u_m/\lambda)]$, for both vortex
and antivortex cycles, where we took account of the fact that the
main contribution to $Q$ comes from the initial acceleration of the
vortex during the time $\sim \tau$ when the field $B(t)$ is close to
$B_v$. Neglecting $\exp(-u_m/\lambda)\ll 1$, we have
    \begin{equation}
    Q=2\omega\phi_0B_v/\pi\mu_0,\qquad \Omega\to 0
    \label{qo}
    \end{equation}

The field region of the single-vortex penetration $B_v < B_0 < B_2$ defined by Eq. (\ref{b2a}) shrinks as the
frequency decreases. In this narrow field region the effect of vortex viscosity can radically change the dependence
of $Q$ on $B_0$ and $\omega$.  Shown in Fig. 5 are the results of  numerical
solution of Eqs. (\ref{dyn1})- (\ref{dyn3}) and (\ref{w}) for different frequencies and the GL parameters $\kappa=\lambda/\xi_s$.
These data are described well by the following formula
    \begin{equation}
    Q=\frac{2\omega\phi_0B_v}{\pi\mu_0}\left(\frac{B_0^2-B_v^2}{B_v^2}\right)^{\omega\tau_2}, \quad \tau_2 =3.98\tau\kappa^{-2/3}
    \label{q2}
    \end{equation}
which reduces to Eq. (\ref{qo}) for $\omega\to 0$. As follows from
Fig. 5 and Eq. (\ref{q2}), the power $Q$ decreases as
$\omega$ increases because of retardation effects due to vortex
viscosity during the short fraction of the rf period in which
$B(t)>B_v$.

\section{Instabilities at high rf fields}

Once the field $B(t)$ exceeds $B_v$, the vortex rapidly accelerates
reaching the maximum velocity $v_m\simeq\lambda/\tau$:
    \begin{equation}
    v_m=\rho_n/2\mu_0\kappa\lambda.
    \label{vm}
    \end{equation}
Eq. (\ref{vm}) gives  $v_m\simeq  30$ km/s for Nb$_3$Sn and
$v_m\simeq 10$ km/s for Nb. Not only are the so-obtained values of
$v_m$ much higher than the velocity of sound, they may even exceed
the critical BCS pairbreaking velocity \cite{bardeen},
    \begin{equation}
    v_\Delta=\frac{\Delta}{mv_F}=\frac{\hbar}{\pi m\xi},
    \label{vc}
    \end{equation}
where $\xi = \hbar v_F/\pi\Delta$, $v_F$ is the Fermi velocity and
$m$ is the electron mass. Indeed, taking  $\xi = 40$nm and the free
electron mass $m$, we obtain $v_\Delta  = $0.8 km/s $<v_m$ for Nb,
and $v_\Delta=10$ km/s $<v_m$ for $\xi=3$nm in Nb$_3$Sn. Here we use
the Bardeen-Stephen model for qualitative estimates only, ignoring
many still not well understood mechanisms essential at low
temperatures, for example the effect of quantized electron states in
the core and the core shrinkage due to the Kramer-Pesch effect
\cite{kp}, resulting in the factor $\sim\ln(T_c/T)$ in the
Bardeen-Stepnen formula \cite{kopnin,lnt1,lnt2}. Yet, for strong rf
fields, $B_0\sim B_c$, the linear viscous drag force derived for
small vortex velocities, becomes inadequate. It was first predicted
theoretically \cite{lo} and observed in many experiments
\cite{inst1,inst2,inst3,inst4,inst5,inst6} that the dependence of
$\eta$ on $v$ at high vortex velocities results in a nonmonotonic
viscous drag force $f_v=v\eta(v)$ and jump-wise instabilities.

\subsection{Instabilities of viscous flux flow}

A nonlinear viscous drug force was first calculated by Larkin
and Ovchinnikov (LO) \cite{lo}, who showed that nonequilibrium effects
in the vortex core decrease the drag coefficient $\eta$ as $v$ increases:
    \begin{equation}
    \eta(v)=\frac{\eta_0}{1+v^2/v_0^2},
    \label{lo}
    \end{equation}
where $\eta_0$ is the Bardeen-Stephen viscosity. The critical velocity $v_0$ in the dirty limit is given by:
    \begin{equation}
    v_0\simeq 0.6\left(\frac{\ell_i v_F}{\tau_\epsilon}\right)^{1/2}\left(1-\frac{T}{T_c}\right)^{1/4}
    \label{v0}
    \end{equation}
Here $\ell_i$ is the mean free path due to impurities, and $\tau_\epsilon(T)$
is the quasiparticle energy relaxation time.  Eq. (\ref{lo})
results in a nonmonotonic dependence of viscous drag force,
$f_v=v\eta(v)$ on the vortex velocity:
    \begin{equation}
    f_v(v)=\frac{\eta_1v}{1+v^2/v_0^2}+\eta_2v,
    \label{fv}
    \end{equation}
Here, following the LO approach, we use two effective
viscosities $\eta_1$ and $\eta_2$, where $\eta_2$ phenomenologically
takes into account the transition to the normal state as $v$ reaches
the pairbreaking velocity $v_\Delta$. For $\eta_1<8\eta_2$, the
force $f_v(v)$ always increases as $v$ increases, but for
$\eta_1>8\eta_2$ the N-shaped dependence $f_v(v)$ develops, as shown
in Fig. 6.  For $\eta_2=0$, and $\eta_1=\eta_0$, the drag force
reaches the maximum value $F_m=\eta_0 v_0/2$ at $v=v_0$.

    \begin{figure}                  %FIG1
    \epsfxsize= 0.80\hsize
    \centerline{
    \vbox{
    \epsffile{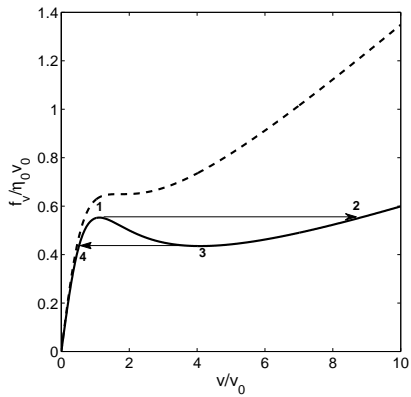}
    }}
    \caption{ The viscous drag force $f_v(v)$ as a function of the vortex velocity
    for the LO instability. The dashed curve shows $f_v(v)$ at $\eta_0=8\eta_i$.
    For $\eta_0>8\eta_i$, the N-shaped dependence $f_v(v)$ develops, as illustrated by the solid curve plotted for
    $\eta_i=0.05\eta_0$. The arrows show the jumpwise change of $v(F)$,
    as the driving force $F$ increases above the maximum value of $f_v(1)$ and then decreases below the minimum value $f_v(3)$.
    }
    \label{Fig.6}
    \end{figure}

The LO instability was originally associated with acceleration of
normal quasiparticles in the vortex core by electric field, which
can increase their energy above $\Delta$. In this case
quasiparticles can escape the core if the diffusion length
$L_D=(D\tau_\epsilon)^{1/2}$ exceeds the core size. The resulting
quasiparticle depletion in the core reduces the core size and the
vortex viscosity according to Eqs. (\ref{lo}) and (\ref{v0}).
However, Eq. (\ref{lo}) is actually more general and may result from
several different mechanisms. In particular, the velocity dependence
(\ref{lo}) can result from coupling of the vortex motion
with other diffusion process, including quasiparticle or temperature
diffusion around the moving vortex core. For example, the electron
overheating of the core can lead to Eq. (\ref{lo}) as follows.

The power $\eta v^2$ generated by the viscous drag increases
the electron temperature in the core $T_m$, reducing the vortex
viscosity $\eta(T_m)\propto B_{c2}(T_m)$. Linearizing
$\eta(T_m)\simeq \eta_0(1-T_m/T_c)$, we write the thermal balance
condition
    \begin{equation}
    (T_m-T_0)g=\eta_0(1-T_m/T_c)v^2,
    \label{tball}
    \end{equation}
where $g\sim \pi k/\ln(L_\theta/L_\xi)$ defines the heat flux from
the core due to the thermal conductivity $k(T)$, where the thermal
length $L_\theta$ is of the order of the film thickness $d$ for
ideal cooling, and $L_\xi\sim \sqrt{\xi\lambda}$ is the length
related to the amplitude of vortex penetration. This logarithmic
factor will be calculated in the next section in more detail; here
we just use $\ln(L_\theta/L_\xi)\sim \ln(d/\sqrt{\xi\lambda})$ for
qualitative estimates. Solving Eq. (\ref{tball}) for $T_m$ results
in the velocity-dependent $\eta(T_0)$ of the LO form:
    \begin{eqnarray}
    \eta = \eta_0(1-T_0/T_c)/(1+\eta_0v^2/g),\qquad
    \label{hlo} \\
    v_0=(g/\eta_0)^{1/2}\sim  [\pi k\rho_n/\phi_0B_{c2}(0)\ln(L_\theta/L_\xi)]^{1/2}
    \label{est}
    \end{eqnarray}
Substituting here the low-T quasiparticle thermal conductivity $k\sim k_n(\Delta/T)^2\exp(-\Delta/T)$,
and using the Wiedemann-Frantz law $k_n \rho_n=(\pi
k_B/e)^2T/3$, and $B_{c2}(0)\sim \phi_0/2\pi\xi_0\ell_i$ and
$\xi_0 \sim \hbar v_F/\Delta$ in the dirty limit, we reduce Eq.
(\ref{est}) to Eq. (\ref{v0}). Here the time constant
    \begin{equation}
    \tau_\epsilon\sim
    \frac{\hbar T}{\Delta^2}\ln\left (\frac{L_\theta}{L_\xi}\right)\exp\left(\frac{\Delta}{T}\right)
    \label{tauu}
    \end{equation}
exhibits the exponential temperature dependence similar to the energy relaxation time $\tau_\epsilon$
between quasiparticles and phonons \cite{tau1} in the LO theory.
However, the exponential dependence of $\tau_\epsilon(T)$ in the thermal model is cut off at
lower T where $k$ is limited by phonons.

To estimate $v_0$, we take
$\rho_n=0.2\mu\Omega$m, $k = 10^{-2}$W/mK, $B_{c2}=23$T for Nb$_3$Sn
at low temperatures \cite{nb3sn} and $L_\xi\sim\sqrt{\lambda\xi}\sim
16$ nm, $L_\theta\sim d\sim 1$ mm, $\ln(L_\theta/L_\xi)\sim 11$. For
these parameters, Eq. (\ref{est}) gives $v_0\sim 0.1$km/s, much
smaller than the estimates for $v_\Delta$ and $v_m$. Thus,
overheating does result in the same Eq. (\ref{lo}),
although in this case vortex core expands as it becomes warmer at
higher velocities \cite{heat1,heat2}, in contrast to the LO core
shrinkage. Moreover, the critical velocity $v_0$ defined by Eq.
(\ref{est}) remains constant at $T_c$, unlike vanishing $v_0(T_0)$
for the LO mechanism, which dominates at $T\approx T_c$. However,
for $T\to T_c$, both Eq. (\ref{v0}) and (\ref{est}) predict the
critical velocity $v_0(T)$ to exceed the linear viscous drag-limited
velocity $v_m(T)\propto (1-T/T_c)^{1/2}$ given by Eq. (\ref{vm}). As
a result, Eq. (\ref{dyn1}) adequately describes rf vortex dynamics
at strong fields $B_0\sim B_c$ and temperatures close to $T_c$.

To evaluate the overheating mechanism in more detail, we assume that
$\eta(T_m)$ depends on a local electron temperature
$T_m(t)$ in the vortex core. The distribution of $T({\bf r},t)$ around a moving vortex is
described by a thermal diffusion equation,
    \begin{equation}
    C\dot{T}=k\nabla^2T - \alpha(T-T_0)+
    \eta(T_m)v^2(t)f[x-u(t),y],
    \label{tdeq}
    \end{equation}
which, after re-definition of the coefficients, can be reduced to
the same mathematical form as the diffusion equation for
nonequilibrium quasiparticles. Here $C$ is the heat capacity, $k$ is
the thermal conductivity, $u(t)$ is the coordinate of the vortex
core moving with the velocity $v(t)$, and the
function $f(x,y)$ accounts for the finite core size, so that $\int
f(r)d^2r=1$. The term $\alpha(T-T_0)$ describes heat exchange with
the environment. For example, $\alpha = h/d$ in a thin film of
thickness $d$ where $h$ is the Kapitza conductance at the sample
surface, and $T_0$ is the bath temperature. For electron overheating, the parameter $\alpha =
C/\tau_\epsilon$ describes heat exchange between electrons and the
lattice, where $C$ is the electron specific heat and $\tau_\epsilon$
is the time of inelastic scattering of quasiparticles in the vortex
core on phonons \cite{heat2,tau1,tau2,tau3}. The last term in the r.h.s. of Eq. (\ref{tdeq})
describes dissipation in the vortex core proportional to the
viscosity $\eta(v,T_m)$ taken at the local core
temperature $T_m(t)$, which, in turn, depends on $v(t)$.
The core form factor $f(r)$ is modeled by the Gaussian function
$f(r)=\pi^{-1}\xi_1^{-2}\exp(-r^2/\xi_1^2)$, where the core radius $\xi_1$ can be smaller
than $\xi$ at $T\ll T_c$ due to the Kramer-Pesch
effect \cite{kp} ( the solutions of Eq. (\ref{tdeq}) depend weakly on  $\xi_1$). We
also consider weak overheating for which dependencies of $C$, $k$ and $\alpha$ on $T$ can be
neglected. As shown below, $T$ can be regarded as either the electron or the lattice temperature,
depending on the time scale of the vortex dynamics involved.

The solution of Eq. (\ref{tdeq}) given in Appendix A, results in the
following integral equation for the temperature $T_m(t)$ in the
vortex core moving with a time-dependent velocity $v(t)$ near the surface:
  \begin{eqnarray}
    T_m(t)=T_0+\frac{1}{\pi k}\int_0^\infty\frac{dt'q(t-t')e^{-t'/t_\theta}}{4t'+t_s}\times \nonumber \\
    \{\exp\left[ -\frac{[u(t)-u(t-t')]^2}{(4t'+t_s)D}\right]\pm
    \nonumber \\
    \exp\left[ -\frac{[u(t)+u(t-t')]^2}{(4t'+t_s)D}\right]\},
    \label{tm}
    \end{eqnarray}
where $q(t)=\eta[T_m(t),v(t)]v^2(t)$ is the time-dependent power
generated by the moving vortex, $D=k/C$ is the thermal diffusivity,
$t_s= \xi_1^2/D$ is the diffusion time across the vortex core, and
$t_\theta=C/\alpha$ is the electron energy relaxation time. The
second term in the parenthesis describes the effect of the surface:
the plus sign corresponds to the thermally-insulated surface,
$\partial_xT(x,t)|_{x=0}=0$, and the minus sign corresponds to the
ideal cooling, $T(0,t)=T_0$. Here we do not consider microscopic
thermal gradients inside the vortex core \cite{clem}, assuming that
their effect is included in the bare $\eta$.

The integral Eq. (\ref{tm}) takes into account retardation effects
due to diffusive redistribution of $T({\bf r},t)$ around an
accelerating vortex, so $T_m(t)$ depends on the vortex velocity
$v(t-t')$ at earlier times. The effect of the surface makes $T_m(t)$
dependent on the vortex coordinate $u(t)$ as well. Eq. (\ref{tm})
simplifies considerably if $v(t)$ varies slowly over the relaxation
time $t_\theta$, and $t_s\ll t_\theta$, and $u(t)>\xi_1$. Then
$q(t-t')$ can be taken out of the integral at $t'=0$, and Eq.
(\ref{tm}) yields the following equation for the local temperature
difference $\delta T_m=T_m-T_0$:
    \begin{equation}
    \delta T_m=\frac{\eta(T_m)v^2}{2\pi k}\left[ \ln\frac{L_\theta}{\tilde{\xi}}
    \pm K_0\left( \frac{2u}{L_\theta} \right) \right]
    \label{hb}
    \end{equation}
Here $\tilde {\xi}=\xi_1e^{\gamma/2}/2\approx 0.67\xi_1$, and
$\gamma = 0.577$ is the Euler constant \cite{abram}. The second term
in the parenthesis decreases exponentially for $u > L_\theta$, and
logarithmically, $K_0(z)\simeq \ln(2/z) -\gamma$ for $\tilde{\xi}\ll
u\ll L_\theta$. In this case the expression in the parenthesis
reduces to $\ln (L_\theta/\tilde{\xi})+\ln (L_\theta/{\tilde u})$,
where $\tilde{u}=ue^{\gamma}/2$. Taking the characteristic amplitude
of vortex penetration $\tilde{u}\sim \lambda$, we can present the
logarithmic part in the form $2\ln(L_\theta/L_\xi)$, where
$L_\xi=(\tilde{u}\tilde{\xi})^{1/2}\sim \sqrt{\xi\lambda}$ was used
before to obtain Eq. (\ref{lo}) in the thermal model. The weak logarithmic
dependence of $v_0$ on $L_\xi$ and $L_\theta$ makes Eq. (\ref{est}) nearly
insensitive to the details of heat transfer and the behavior of $u(t)$.

    \begin{figure}                  %FIG1
    \epsfxsize= 0.75\hsize
    \centerline{
    \vbox{
    \epsffile{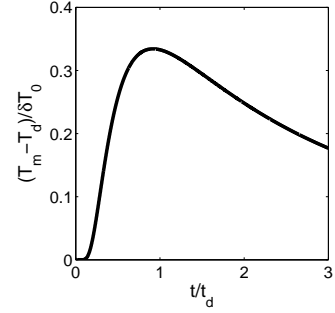}
    }}
    \caption{Temperature spike in the vortex core after the jump described by Eq. (\ref{spike})
    for $t_\theta=10 t_d$, $t_d=\delta u^2/4D$, $\delta T_0=WD/\pi k\delta
    u^2$, and $T_d=T_0+\eta_0v_0^2\ln(L_\theta/\tilde{\xi})/2\pi k$.
    }
    \label{Fig.7}
    \end{figure}

In the other limiting case of very rapid variation of $v(t)$, the
vortex reaches the critical velocity $v_0$ and then jumps by the
distance $\delta u$, dissipating the energy $W$ during the short time $\delta t$. Then
$q(t)=q_0+W\delta(t)$ and Eq. (\ref{tm}) results in the
following implicit equation for $\delta T_m(t)$ at $t>\delta t$:
    \begin{equation}
     \delta T_m(t)\simeq \frac{\eta_0v_0^2}{2\pi k}\ln\frac{L_\theta}{\tilde{\xi}}
     +\frac{W(T_m)}{4\pi kt}e^{-\delta u^2/4Dt-t/t_\theta},
    \label{spike}
    \end{equation}
which describes a temperature spike in the core followed by
relaxation of $\delta T_m(t)$, as shown in Fig. 7. Here the first
term in the r.h.s. gives $\delta T_m$ before the jump, and the
effect of the surface is neglected.

    \begin{figure}                  %FIG1
    \epsfxsize= 0.85\hsize
    \centerline{
    \vbox{
    \epsffile{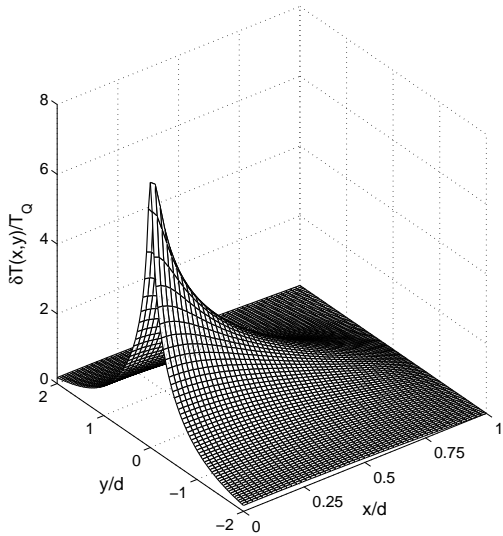}
    }}
    \caption{ Temperature variation $\delta T(x,y)=T(x,y)-T_0$ around a heat source given by Eqs. (\ref{txy}) and (\ref{tx}) for
    $u_m=10^{-3}d/\pi$ and normalized by $T_Q=Q/2\pi k$.
    }
    \label{Fig.8}
    \end{figure}

Next we consider the steady-state temperature field $T({\bf r})$
averaged over rf oscillations, where $T({\bf r})$ is determined by
the balance of the vortex heat source and thermal diffusion.
Solution of the thermal diffusion equation in Appendix A yields the
following distribution of $\delta T({\bf r})=T({\bf r})-T_0$ from a
heat source localized at the thermally-insulated surface $(x=0)$ of
a slab of thickness $d$, ideally cooled from the other side, $\delta
T(d)=0$:
    \begin{equation}
    \delta T({\bf r})=
    \frac{1}{2\pi k}\int_0^dq(x')\ln\frac{\cosh\frac{\pi y}{2d}+
    \cos\frac{\pi(x+x')}{2d}}{\cosh\frac{\pi y}{2d}-\cos\frac{\pi(x-x')}{2d}}dx'
    \label{txy}
    \end{equation}
Here $q(x)$ is the power density averaged over the rf period.
On scales greater than the size of the heat source, $\delta T({\bf r})$
depends only on the total power $Q=\int_0^dq(x)dx$:
    \begin{equation}
   \delta T({\bf r})=\frac{Q}{2\pi k}\ln\frac{\cosh(\pi y/2d)+\cos(\pi x/2d)}{\cosh(\pi y/2d)-\cos(\pi x/2d)},
    \label{tx}
    \end{equation}
Here $\delta T({\bf r})$ decays exponentially over the length
$2d/\pi$ as shown in Fig. 8. Near the heat source $\delta T({\bf r})$
weakly (logarithmically) depends on details of $q(x)$. As shown in Appendix A,  the distribution of
$\delta T(0,y)$ along the surface to the logarithmic accuracy is
given by
    \begin{equation}
    \delta T(0,y)=\frac{Q}{2\pi k}\ln\frac{16d^2}{\pi^2(y^2+r_0^2)},\qquad y^2\ll d^2,
    \label{dty}
    \end{equation}
where $r_0$ quantifies a size of dissipation source. The account of finite $r_0$
cuts off the logarithmic divergence in Eq. (\ref{tx}) at $x=y=0$, resulting in a maximum
temperature disturbance at $y=0$:
    \begin{equation}
    \delta T_m\simeq \frac{Q}{\pi k}\ln\frac{4d}{\pi r_0}
    \label{tmm}
    \end{equation}
Eq. (\ref{tmm}) reduces to Eq. (\ref{hb}) with $u\sim r_0$ and
$L_\theta\sim d$.

The physical meaning of $T$ in the above formulas depends on the
relevant vortex time scales. For example, for the supersound vortex
penetration or vortex jumps on the time scale much shorter than the
electron-phonon energy relaxation time, the quasiparticle are not in
equilibrium with the lattice, and $T({\bf r},t)$ in Eq. (\ref{spike})
can be regarded as an electron core temperature. However, steady-state
vortex oscillations in the rf field generate a dc power, which must be
transferred to the coolant through phonons. In this case Eqs.
(\ref{txy}) describes the lattice temperature distribution around a
vortex if the phonon mean free path is shorter than the film
thickness \cite{bs}. Thus, the vortex oscillates in a "warm tunnel"
with the lattice temperature $\delta T({\bf r})$ shown in Fig. 8, but in
addition to that the vortex core gets overheated with respect to the
lattice during short periods of rapid acceleration, jumps or
annihilation with antivortices, as described before.

\subsection{Jumpwise vortex penetration}

For the LO vortex drag coefficient $\eta(v)$ given by Eq. (\ref{lo}), the equation of motion becomes
    \begin{eqnarray}
    \frac{\eta_0\dot{u}}{1+\dot{u}^2/v_0^2}=\frac{\phi_0B_0}{\mu_0\lambda}e^{-u/\lambda}\sin\omega t-
    \nonumber \\
    \frac{\phi_0^2}{2\pi\mu_0\lambda^3}
    K_1\left[\frac{2}{\lambda}\sqrt{u^2+\xi_s^2}\right].
    \label{dyni}
    \end{eqnarray}
The nonmonotonic velocity dependence of the viscous drag force in the l.h.s. of  Eq.
(\ref{dyni}) qualitatively changes vortex dynamics as $v(t)$ exceeds
the critical value $v_0$ for which the viscous force reaches the
maximum $F_m=\eta_0 v_0/2$. Indeed, the differential equation for
$u(t)$ has the form $\eta_0\dot{u}/(1+{\dot u}^2/v_0^2)=F(u,t)$,
where $F$ is the net electromagnetic force given by the r.h.s. of Eq. (\ref{dyni}). We
can introduce the ratio $P$ of the maximum Lorentz driving force at
$B_0=B_v$ to the maximum viscous force:
    \begin{equation}
    P=\frac{2\phi_0B_v}{\mu_0\lambda\eta_0v_0},
    \label{p}
    \end{equation}
As shown above, the
Bardeen-Stephen viscous flow results in unphysically high vortex
velocities, indicating that $P\gg 1$ and the dependence of $\eta$ on $v$ must be
taken into account. However, in this case there are
regions at the surface where $F(u,t)$ exceeds $F_m$ as shown in Fig. 9.
In these regions the force balance Eq. (\ref{dyni}) cannot be satisfied
and the vortex jumps to the place where $F(x)\leq F_m$.
To see how it happens, we present the quadratic equation Eq. (\ref{dyni}) for
$\dot{u}$ in the form
    \begin{equation}
    \dot{u}=v_\pm(F)=\frac{v_0F_m}{F(u,t)}\left[1\pm\sqrt{1-\frac{F^2(u,t)}{F_m^2}}\right],
    \label{qeq}
    \end{equation}
where $F_m=\eta_0v_0/2$. For $v\ll v_0$ and $F\ll F_m$, Eq.
(\ref{qeq}) with the minus sign in the brackets reduces to Eq. (\ref{dyn1}).

    \begin{figure}                  %FIG1
    \epsfxsize= 0.75\hsize
    \centerline{
    \vbox{
    \epsffile{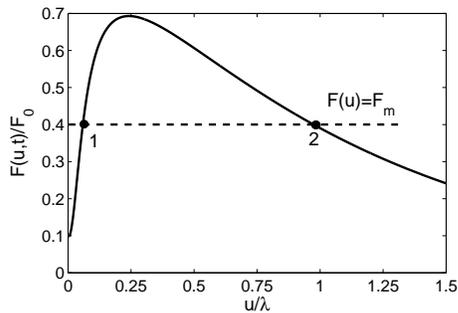}
    }}
    \caption{ Instantaneous profile of the force $F(u)$ normalized by
    $F_0=\phi_0B_v/\mu_0\lambda$ for $B(t)=1.1B_v$ and $\kappa=20$.
    The dashed line shows the maximum viscous force
    $F_m=\eta_0v_0/2$, which can balance $F(u)$ only in the
    regions where $F(u)<F_m$. Thus, vortex first moves from
    $u=0$ to $u=u_1$, then jumps from point 1 to
    point 2 after which it moves continuously as described by Eq. (\ref{dyni}) until
    the next jump and annihilation with the antivortex on the way back.
    }
    \label{Fig.9}
    \end{figure}

Penetration of the vortex at $B(t)>B_v$ is therefore described by
the first order differential equation $\dot{u}=v_-(F)$, which is
well defined only if $F(u,t)\leq F_m$, otherwise the driving force
exceeds the maximum friction force, and the square root in Eq.
(\ref{qeq}) becomes imaginary. Vortex dynamics in this case can be
understood from Fig. 9, which shows an instantaneous profile of
$F(u,t)$ for $B(t)>B_v$. Here a vortex enters the sample with zero
velocity at $t=t_0$ and then accelerates because the net force
$F(u,t)$ increases as the vortex moves away from the surface and the
vortex-image attraction weakens. This part of $u(t)$ is described by
the equation $\dot{u}=v_-(F)$ until the vortex reaches the point 1
where $F(u,t)=F_m$. In the region $u>u_1$, the friction force cannot
balance the driving force, so the vortex jumps to a point 2 where
$F(u,t)=F_m$ and the viscous drag can balance the driving force. As
follows from Eq. (\ref{spike}), the core temperature does not change
right after the jump $(t\to 0)$, but then starts increasing. After
that the smooth parts of $u(t)$ are described by to the same dynamic
equation $\dot{u}=v_-(F)$ for the vortex, which reaches the maximum
penetration depth $u_m$ then turns around and accelerates toward the
surface during the negative rf half period. However, as the vortex
reaches the negative critical velocity $-v_0$ on the way out, it
jumps again and either exits the sample or collides with the
incoming antivortex and annihilates as described in the previous
section. The positions of the jumps $u_j$ at the corresponding times
$t=t_j$ are determined by the equation $F(u_j,t_j)=\pm F_m$:
    \begin{equation}
    \pm\frac{\eta_0v_0}{2}=\frac{\phi_0B_0e^{-u_j/\lambda}}{\mu_0\lambda}\sin\omega t_j-
    \frac{\phi_0^2}{2\pi\mu_0\lambda^3}
    K_1\!\left(\frac{2u_j}{\lambda}\right),
    \label{jump1}
    \end{equation}
where the $\pm$ signs should be taken for the incoming and outgoing
parts of $u(t)$, respectively.
    \begin{figure}                  %FIG1
    \epsfxsize= 0.7\hsize
    \centerline{
    \vbox{
    \epsffile{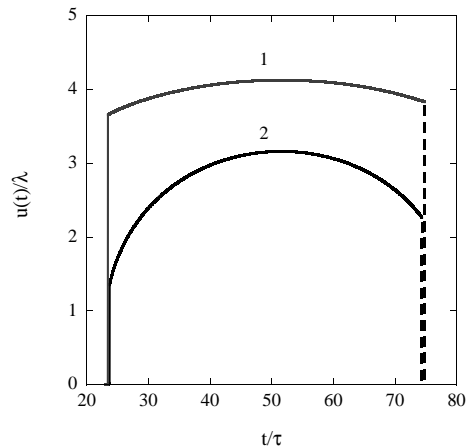}
    }}
    \caption{Jumpwise vortex penetration calculated by solving Eq. (\ref{dyni})
    for $\kappa=34$, $\omega\tau = 0.244$, $B_0=1.02B_v$ and: $v_0= 0.052v_m$, $P=38.4$ (1),
    and $v_0 = 0.52v_m$, $P=3.84$ (2). The dashed lines show the antivortex penetration.
    }
    \label{Fig.10}
    \end{figure}
Shown in Fig. 10, are examples of vortex penetration dynamics calculated numerically
from Eq. (\ref{dyni}). In this case the antivortex jumps in and annihilates with the
outgoing vortex before the vortex reaches the critical velocity $-v_0$.

As mentioned above, the LO instability facilitates a quick evolution
of the vortex semi-loop originating at a 3D surface defect into a
straight vortex parallel to the surface. Indeed, as evident from
Fig. 2, the vortex propagation velocity is maximum for the segments
of the semi-loop perpendicular to the surface because they are
driven by the continuously increasing maximum Lorentz force
$F_L=B(t)\phi_0/\mu_0\lambda$, while experiencing no
counterbalancing image forces. As a result, the LO instability first
occurs for the perpendicular vortex segments, causing them to jump
along the surface to the place where the viscous force is able to
balance the Lorentz force. However, unlike the parallel vortex
segments whose jump distance $\sim\lambda$ perpendicular to the
surface is limited by the London screening, the jump length along
the surface of a semi-infinite sample is infinite because $F_L$
remains constant. Thus, the vortex semi-loop turns into a straight
vortex in a jumpwise manner when the lateral velocity of the
perpendicular vortex segments reach $\pm v_0$.

Several points should be made regarding the jumpwise vortex
dynamics. First, the vortex trajectory $u(t)$ comprised of the jumps
connected by smooth parts described by the equation $\dot{u}=v_-(F)$
occurs only if $\dot{F}(u_j,t_j)<0$ at the jump points where
$v(t_j)=\pm v_0$. However, at higher frequencies, or as the vortex
overheating is taken into account, there are situations when
$\dot{F}(u_j,t_j)>0$. In this case the vortex velocity after the
jump exceeds $v_0$ and the smooth parts of $u(t)$ are described by
both branches $\dot{u}=v_\pm(F)$, as shown in Appendix B.

The second point is that, for the overdamped dynamics described by
Eq. (\ref{dyni}), the jumps occur instantaneously unless the second
ascending branch due to $\eta_2$ term in Eq. (\ref{fv}) is taken
into account. However, this branch in the LO model corresponds to
very high velocities $\sim v_\Delta$, for which the adequate theory
of the nonequilibrium vortex core structure and the vortex drag
force is lacking. In our phenomenological London approach we assumed
that the vortex jumps to the nearest point $u_2$ where the friction
force is able to balance the driving force $F(u_2)=F_m$. However,
the instantaneous LO dependence $\eta(v)$ does not include
retardation effects due to finite relaxation times of the
superconducting order parameter, or diffusive redistribution of
nonequilibrium quasiparticles or temperature around a rapidly
accelerating/decelerating vortex core. These effects are taken into
account by the integral Eq.(\ref{tm}) for the core temperature
$T_m(t)$, which shows that the vortex jump time and length are
affected by intrinsic dynamics of $\eta$. Thus, there is a diffusion
time scale $\delta t\sim \delta u^2/D$ for the vortex jump by the
distance $\delta u$, where $D$ equals either $k/C$ in the thermal
model or the quasiparticle diffusivity in the LO theory. In the
thermal model this estimate gives $\delta t\sim 4\times 10^{-12}$ s
if we take $\delta u\simeq\lambda=65$nm, $k\simeq 0.1$W/mK and
$C\sim 100$J/m$^3$K for Nb$_3$Sn at 2K \cite{nb3sn}, or even much
shorter time for Nb, for which $\lambda=40$nm and $\kappa\simeq
10$W/mK. The so-estimated $\delta t$ is smaller than the inverse gap
frequency, indicating that once the overheated core gets in the
region where the friction force is able to balance the driving
force, it cools down very quickly due to the electron component of
thermal conductivity. In Nb the electron thermal conductivity
$k\propto\exp (-\Delta/T)$ is still significant down to $T>1K$, but
at lower temperatures $\delta t$ may be limited by much slower
phonon irradiation from the overheated core. At the same time, the
electron temperature relaxation time $\tau_\epsilon$ outside the
core results from a slow phonon-mediated recombination of
quasiparticles \cite{tau1}
    \begin{equation}
    \tau_\epsilon\simeq \tau_0
    \left(\frac{\Delta}{T}\right)^{1/2}\exp\left(\frac{\Delta}{T}\right),
    \label{tef}
    \end{equation}
which yields $\tau_\epsilon\sim 30$ ns much longer than the thermal
diffusion time $\delta t$ for Nb at 2K. For $\omega \tau_\epsilon\ll
1$, the quasiparticles are overheated with respect to the lattice,
in a highly inhomogeneous way according to the distribution of the
lattice temperature $T({\bf r})$ shown in Fig. 8. In this case the
condition $\omega \tau_\epsilon({\bf r})> 1$ of the electron overheating
can locally be satisfied in colder regions away from the core, but
near the vortex core the electron temperature can be close to the
lattice temperature if $\omega \tau_\epsilon({\bf r}) <1$ because of higher
$T({\bf r})$, which greatly accelerates the energy exchange between electron and
phonons. For example, for Nb at $T_0=2K$, the local
increase of the lattice temperature to $T=4K$ yields $\tau_\epsilon
(T)\simeq \tau_\epsilon(T_0)\exp(-\Delta/T_0+\Delta/T)$, giving
$\tau_\epsilon\sim 0.3$ ns, and $\omega \tau_\epsilon\sim 1$ at $1-2$ GHz.

Another contribution to $\delta t$ comes from a finite vortex mass
$M$. In the Suhl model $M=2mk_F/\pi^3$ is due to localized electrons
in the vortex core, where $k_F=(3\pi^2n)^{1/3}$ is the Fermi wave
vector \cite{suhl}. The jump by $\lambda$ due to the driving force
$F=\phi_0B_v/\mu_0\lambda$ takes the time $\delta t$ set by the
Newton law $2\lambda/\delta t^2\simeq F$. Using
$B_v=\phi_0/4\pi\lambda\xi$, $\lambda^2=\mu_0m/ne^2$, $\xi=\hbar
v_F/\pi\Delta$, we obtain
    \begin{equation}
    \delta t\simeq \frac{4\hbar}{\Delta}\kappa^{1/2},
    \label{dt}
    \end{equation}
A more accurate account of quantized levels in the vortex core
\cite{vm1,vm2} or lattice deformation around the moving vortex
\cite{vm3,vm4,vm5} can increase the vortex mass, thus further
increasing the jump time $\delta t$. Yet, although vortex jumps are
quantified by the multiple relaxation times discussed above, they
seem to occur much faster than the rf periods we are dealing with in
this work.

\subsection{Rf dissipation}

Power $Q$ dissipated with the account
of the jumpwise vortex instabilities can be written in the form
    \begin{eqnarray}
    Q=\frac{\omega}{\pi}\bigl{\{}\oint \eta(\dot{u}){\dot u}^2dt+\oint\eta(\dot{u})({\dot u_1}^2+{\dot u_2}^2)dt+ \nonumber \\
    \sum_m[G(t_m,u_{m+})-G(t_m,u_{m-})]\bigr{\}},
    \label{wj}
    \end{eqnarray}
where the integrals are taken over all smooth parts of the vortex trajectory $u(t)$,
including the vortex-antivortex annihilation parts, like in Eq. (\ref{w}). The last term in Eq. (\ref{wj})
is the sum of energies released during all vortex jumps at $t=t_m$, from $u=u_{m-}$ to $u=u_{m+}$, where
the instantaneous free energy $G(u,t)$ is given by Eq. (\ref{gu}). Shown in Fig. 11 is the second vortex penetration field $B_2$
calculated by solving Eqs. (\ref{dyni}), (\ref{jump1}), and
(\ref{wj}) numerically for the parameters of Nb$_3$Sn. Here $B_2$
cannot be fit with a power-law similar to Eq. (\ref{b2a}).

The LO instability makes the behavior of $Q(B_0,\omega)$ more
complicated as compared to $Q(B_0)$ described by Eq. (\ref{q2}).
As shown in Fig. 12, there are
three distinct field regions of very different vortex dynamics. The
pure Bardeen-Stephen dynamics like that shown in Fig. 3 is limited
to a very narrow region of $B_0$ close to $B_v$ (labeled by $a$ in the inset).
In this case the vortex penetration depth $u_m$ turns out
to be smaller than $\xi$, indicating that the London theory combined
with the Bardeen-Stephen drag cannot give a self-consistent
description of vortex dynamics at low temperatures. However, Eq.
(\ref{dyn1}) adequately describes rf vortex dynamics at
higher $T$ close to $T_c$ where the LO instability is irrelevant
because the critical velocity $v_0\propto (1 - T/T_c)^{1/4}$ becomes
larger than $v_m\propto (1 - T/T_c)^{1/2}$.

\begin{figure}                  %FIG1
    \epsfxsize= 0.6\hsize
    \centerline{
    \vbox{
    \epsffile{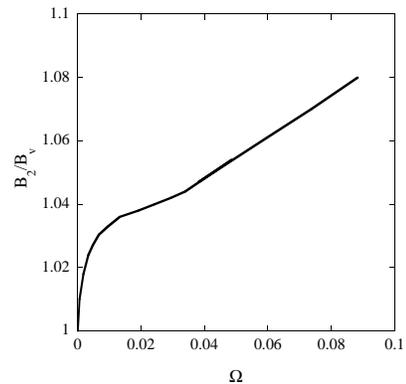}
    }}
    \caption{The dependence of the second penetration field $B_2$ on the dimensionless frequency $\Omega=\omega\tau/\kappa$
    for the LO instability at $\kappa=34$ and $v_0=0.52v_m$.
    }
    \label{Fig.11}
    \end{figure}

 \begin{figure}                  %FIG1
    \epsfxsize= 0.75\hsize
    \centerline{
    \vbox{
    \epsffile{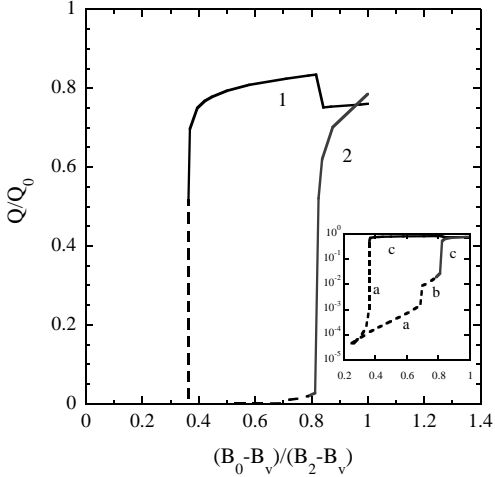}
    }}
    \caption{$Q(B_0)$ calculated with the account of the LO instability for
    $\kappa=34$, $v_0=0.52v_m$, and $\Omega = 0.019$ (1) and $\Omega = 0.089$
    (2). Solid lines show the parts of $Q(B_0)$, for which
    $u_m>\xi_s$. Inset shows the detailed behavior of $Q(B_0)$ close to $B_v$, where the parts labeled $a$
    correspond to the pure Bardeen-Stephen dynamics for the entire $u(t)$. Parts b
    are partially controlled by the Bardeen-Stephen drag for the vortex reaching $u_m$,
    then turning back and then jumping to the surface
    from $u=u_c<u_m$ due to the LO instability. Parts c correspond to
    the LO dynamics similar to that in Fig. 10.
    }
    \label{Fig.12}
    \end{figure}

The parts of the $Q(B_0)$ curves labeled $b$ in Fig. 12 correspond
to an intermediate case, for which the most part of $u(t)$,
including the initial acceleration of the vortex, reaching the
maximum penetration depth $u_m$, and turning back, does not involve
the LO instability. However, as the velocity of the exiting vortex
exceeds $-v_0$, it jumps to the surface and disappears,
significantly increasing the dissipated power $Q$. Further increase
of $B_0$ corresponds to the parts of $Q(B_0)$ curves labeled $c$,
for which the LO instabilities occur both on the penetration and the
exit parts of $u(t)$, like those in Fig. 10. In this case $Q(B_0)$
jumps up to a much higher level $Q\sim Q_0$ until the second
penetration field $B_2$ is reached. Here the behavior of $Q(B_0)$
can also depend on $\omega$: for lower frequency (curve 1), $Q(B_0)$
increases weakly between $0.4 < \varepsilon(B_0) <0.8$, but for
$\varepsilon =(B_0-B_v)/(B_2-B_v)>0.8$, the power $Q(B_0)$ jumps
down. This behavior reflects the change in the vortex dynamics: for
$0.4<\varepsilon <0.8$, the vortex jumps out of the
sample before the antivortex enters, while for $\epsilon>0.8$, the jump of the vortex
toward the surface is accompanied by the penetration of the antivortex and their
annihilation, like that in Fig. 10.  For higher frequencies, this
change in the vortex dynamics formally occurs only for $B_0>B_2$, so
the down step in $Q(B_0)$ does not show up in curve 2 in Fig. 12.

\subsection{Thermal self-localization of vortex penetration}

Local temperature increase around oscillating vortex reduces both
critical fields $\tilde{B}_v=B_v(T_m)$ and $\tilde{B}_2=B_2(T_m)$ as
compared to their isothermal values $B_v(T_0)$ and $B_2(T_0)$. To
evaluate this effect we combine Eq. (\ref{q2}) for $Q$ and Eq.
(\ref{tmm}) for $T_m$ and obtain for $\omega\tau_2\ll 1$:
  \begin{equation}
   \delta T_m\simeq \frac{2\omega\phi_0B_v}{\pi^2\mu_0k\ln(4d/\pi r_0)},
  \label{gg}
    \end{equation}
Next we linearize $\tilde{B}_v\simeq B_v-|B_v'|\delta T_m$ with
respect to $\delta T_m$, where $B_v'=\partial_TB_v(T_0)$. Then the
effective field $\tilde{B}_v$ and $\tilde{B}_2$ take the form:
    \begin{eqnarray}
    \tilde{B}_v=(1-b)B_v,
    \qquad
    \tilde{B}_2=(1-b)B_2,
    \label{b2v} \\
    b=\frac{2\omega\phi_0|B_v'|}{\pi^2\mu_0k\ln(4d/\pi r_0)}\qquad
    \label{b}
    \end{eqnarray}
Thus, both $B_2$ and $B_v$ are reduced by dissipation, which only
produces constant shifts of local $B_v$ values but does not change
their initial distribution. For Nb, taking $k=10$ W/mK, $B_v'\simeq
B_v/T_c$, $B_v=0.15$ T, $\omega/2\pi = 2$ GHz, $\ln (4d/\pi
r_0)\simeq 10$, as before, we obtain a small value $b\sim 3\times
10^{-4}$, for which the shift of $B_v$ is negligible. However, for
Nb$_3$Sn, with $\kappa\sim 10^{-2}$ W/mK, we get a much higher value
$b\sim 0.3$, indicating that dissipation can significantly reduce
$B_v$ and $B_2$, expanding the field region $\tilde{B}_v<B_0<B_s$ of
individual vortex penetration.

The condition of the single-vortex penetration $B_v<B_0<B_2$ implies
that the local value of $B_2$ is smaller than the uniform
superheating field $B_{s}$. Multiple vortex penetration for
$B_0>B_2$ causes strong dissipation, further decreasing $B_2$
and resulting in avalanche-type dendritic vortex penetration
\cite{den1,den2}. Such thermo-magnetic dendritic flux avalanches
have been observed in both low-$T_c$ and high-$T_c$ superconductors
\cite{denYBCO,denNb1,denNb2,denNb3Sn,denMgB2,denNi}. Notice that the
superfast vortex penetration through the surface barrier due to the
jumpwise LO vortex instability may pertain to the supersonic vortex
velocities observed for dendritic vortex penetration in
YBa$_2$Cu$_3$O$_7$ and YNi$_2$B$_2$C films \cite{denYBCO,denNi}.

Temperature distribution (\ref{tx}) results in the long-range dc
repulsion force ${\bf f}_T(L)=-s^*\nabla T$ between two oscillating
vortices spaced by $L$. Here $s^*(T)$ is the vortex transport entropy
responsible for thermomagnetic effects in the mixed state
\cite{s1,s2}. In thick films $d\gg \lambda$, vortices are localized
at the surface, so to calculate the thermal force $f_T(L)$, we put
$x=0$ in Eq. (\ref{tx}), and obtain
    \begin{equation}
    f_T(L)=\frac{s^*Q}{2dk\sinh(\pi L/2d)}
    \label{ft}
    \end{equation}
This long-range force on the scale much greater than $\lambda$
results in repulsion of neighboring vortex penetration channels,
facilitating bending instability and dendritic branching of the
multi-vortex tracks.

\section{Effect of pinning}

\subsection{Trapping rf vortices at strong fields, $B_0\sim B_v$}

At the initial stage of vortex penetration, $B(t)\simeq B_v$,
the driving force $F_L$ is much stronger than typical pinning forces by
materials defects. However, as the vortex moves deeper in the
sample, the force $F_L(u)\propto \exp(-u/\lambda)\sin\omega t$
decreases exponentially, so pinning becomes more effective if
the vortex trajectory passes a pin aligned with the place of the vortex entry
within a "belt" $\sim\lambda$ wide. In this case the vortex can
be trapped and stay pinned as the rf field changes sign. However, as $B(t)$ reaches
$-B_v$ during the negative rf cycle, an antivortex penetrates
along the same trajectory as the vortex did and annihilates with the
pinned vortex.

The power dissipated due to the vortex-antivortex annihilation can
be evaluated from the change of the thermodynamic potential $G(u)$:
    \begin{equation}
    Q\simeq \frac{\omega}{\pi}\left[\frac{2B_v}{\phi_0}(1-e^{-u_p/\lambda})+U_0\right]
    \label{qpin}
    \end{equation}
This quasi-static expression differs from Eq. (\ref{qo}) by the
factor $1-\exp(-u_p/\lambda)$, which accounts for the finite
Meissner currents at the pin, $x=u_p$, and the
term due to the gain in the pinning energy $U_0$ as the pinned
vortex annihilates with the antivortex.

\subsection{Residual surface resistance at $B_0\ll B_v$}

    \begin{figure}                  %FIG1
    \epsfxsize= 0.3\hsize
    \centerline{
    \vbox{
    \epsffile{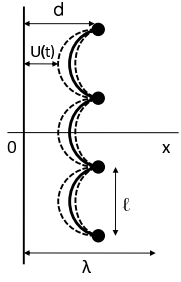}
    }}
    \caption{Vortex pinned by a chain of defects near the surface. The solid line shows the equilibrium vortex shape
    due to competition of pinning and the image attraction forces. The dashed lines show instantaneous
    vortex profiles for $B(t) = B_0$ and $B(t)=-B_0$, between which the vortex line oscillates.
    }
    \label{Fig.13}
    \end{figure}

Besides dissipation due to vortex penetration and exit, there is also dissipation
due to rf oscillations of vortices already trapped in the sample
\cite{rabin}. In this section we address rf dissipation and
depinning of vortices trapped in a superconductor during field
cooling, in the limit of very low flux density, for which the intervortex interaction is negligible.
We first consider a single vortex pinned by a chain
of equidistant defects spaced by $\ell$ from each other and by
$d$ from the surface in the layer of rf field penetration, as
shown in Fig. 13. In this case the equation of motion of the vortex line
takes the form
    \begin{eqnarray}
    \eta_0 \dot{u}=\epsilon u^{''}+\frac{\phi_0B_0}{\mu_0\lambda}e^{-u/\lambda}\sin\omega t- \nonumber \\
    \frac{\phi_0^2}{2\pi\mu_0\lambda^3}
    K_1\left(\frac{2u}{\lambda}\right) +\sum_mf_p(u, y-m\ell),
    \label{dynp}
    \end{eqnarray}
where the first term in the r.h.s describes bending stress of the vortex line, the prime
means differentiation over the coordinate $y$ along the surface, and the last term is the sum of the
elementary pinning forces $f_p(x,y)$. For $\ell>\lambda$, the
dispersive vortex line tension $\epsilon$ reduces to the vortex self
energy $\epsilon = \phi_0B_{c1}/\mu_0$ per unit length \cite{ehb}.

As evident from Fig. 13, the magnetic attraction to the surface makes the pinned vortex not
straight even at zero rf field.  As a result, there is a minimum trapping distance $d_m$, so that only vortices spaced by $u>d_m$
can be pinned. Vortices spaced by $u<d_m$ are unstable and annihilate at the surface, since the image attraction prevails over pinning.
For weak identical pins, $d_m$ can be evaluated from the force balance equation:
   \begin{equation}
    \frac{\phi_0^2}{2\pi\mu_0\lambda^3}
    K_1\left(\frac{2d_m}{\lambda}\right) = \frac{f_p}{\ell},
    \label{dm}
    \end{equation}
where $f_p$ is the maximum elementary pinning force. The vortex segments between the pins bow out toward the surface,
but for vortices in the trapped flux zone $x>d_m$, the curvature of $u(y)$ is weak, and the image attraction force $F_i=(\phi_0^2/2\pi\mu_0\lambda^3)K_1(2u/\lambda)$ is nearly uniform. In this case the equilibrium shape of the vortex segment
between the pins is determined by the equation $\epsilon u_0^{''}=F_i$ with $u_0(\pm \ell/2)=d$, which yields the parabolic profile:
    \begin{eqnarray}
    u_0(y)=d- u_{0m}\left(1-\frac{4y^2}{\ell^2}\right), \qquad
    \label{uu} \\
    u_{0m}=\frac{\phi_0^2\ell^2}{16\pi\mu_0\epsilon \lambda^3}
    K_1\left(\frac{2d_m}{\lambda}\right)= \frac{\ell^2K_1(2d_m/\lambda)}{8\lambda\ln\tilde{\kappa}}.
    \label{uoo}
    \end{eqnarray}
Here $\epsilon=\phi_0^2(\ln\kappa + c_v)/2\pi\mu_0\lambda^2$ where the constant $c_v\approx 0.5$
accounts for the vortex core contribution. It is convenient to use
the effective $\tilde{\kappa}\approx 1.65\kappa$ defined by $\ln{\tilde\kappa}=\ln\kappa + c_v$.
Eqs. (\ref{uu}) and (\ref{uoo}) correspond to $\ell\gg\lambda$, so the condition that $u_{0m}\ll\lambda$ is provided by
$d_m>\lambda$, as follows from Eq. (\ref{dm}). For denser pins, $\ell < \lambda$, the nonlocal expression for
$\epsilon$ should be used \cite{ehb}, in which case $\ln\kappa$ in Eq. (\ref{uoo}) is to be replaced
by $\ln (\ell/\xi)$.

To calculate the power $Q_v$ dissipated by the pinned vortex under the weak
($B_0\ll B_v$) rf field, we seek the solution of Eq. (\ref{dynp}) in
the form $u(y,t)=u_0(y)+\delta u(y,t)$. Here the Fourier component
$\delta u_\omega (y) = \int \delta u(y,t)e^{-i\omega t}dt$ of the
oscillating vortex displacement $\delta u(y,t)$ satisfies the
linearized equation
    \begin{equation}
   i\omega \eta_0\delta u_\omega=\epsilon\delta u_\omega^{''}+f_\omega,
    \label{anz}
    \end{equation}
where $f_\omega=\phi_0B_0\exp(-d/\lambda)/\mu_0\lambda$. In Eq. (\ref{anz}) we neglect the
image force $\propto \exp (-2d/\lambda)\delta u$, which is much smaller than the first
elastic term in the r.h.s. for weak pinning and $d>d_m$ defined by Eq.
(\ref{dm}). The solution of Eq. (\ref{anz}) with the
boundary condition of the fixed vortex ends at the pins, $\delta
u_\omega(\pm \ell/2)$, has the form
    \begin{eqnarray}
    \delta u_\omega = \frac{f_\omega}{i\omega\eta_0}\left(1-\frac{\cos[(1-i)\Omega_p^{1/2}y/\ell]}{\cos[(1-i)\Omega_p^{1/2}/2]}\right),
    \label{uom} \\
     \Omega_p=\omega\tau_p, \qquad \tau_p=\eta_0\ell^2/2\epsilon.\qquad
    \label{taup}
    \end{eqnarray}
Here we introduced the pinning relaxation time constant $\tau_p$ and the dimensionless
frequency $\Omega_p$. The dissipated power per unit vortex length,
    \begin{equation}
    Q_v=\frac{\eta_0\omega^2}{2\ell}\int_{-\ell/2}^{\ell/2}|\delta u_\omega(y)|^2dy
    \label{qom}
    \end{equation}
can be calculated substituting here Eq. (\ref{uom}). A straightforward integration
then yields
    \begin{eqnarray}
    Q_v=\frac{f_\omega^2}{2\eta_0}\Gamma_\omega(\sqrt{\Omega_p}),
    \label{qfin} \\
    \Gamma_\omega(z)=1-\frac{\sinh z+\sin z}
    {z\bigl(\cosh z+\cos z\bigr)}.
    \label{ga}
    \end{eqnarray}
For $\omega\tau_p\gg 1$, Eq. (\ref{qfin}) gives the
frequency-independent limit, $Q_{vm}\to f_\omega^2/2\eta_0$ inversely
proportional to $\eta_0$. However, for $\omega\tau_p\ll
1$, we obtain the quadratic frequency dependence,
$Q_v=f_\omega^2\eta_0\ell^4\omega^2/240\epsilon^2$, proportional to
$\eta_0$.

    \begin{figure}                  %FIG1
    \epsfxsize= 0.8\hsize
    \centerline{
    \vbox{
    \epsffile{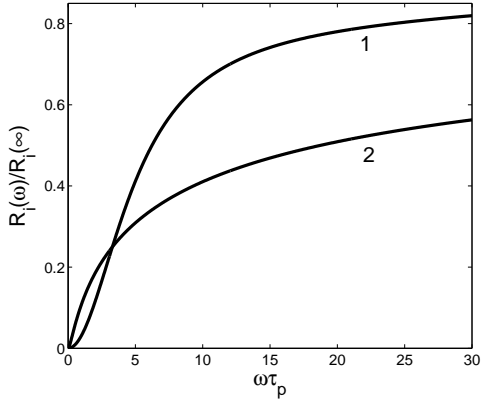}
    }}
    \caption{Frequency dependencies of the residual surface resistance due to pinned vortices. Curve 1 corresponds to $R_i$ for a
    periodic chain of pins described by Eq. (\ref{ri}), and curve 2 corresponds to $R_i$ for the exponential distribution of pinned vortex
    segments given by Eq. (\ref{rif}).  Here $R_i(\infty)=\phi_0^2\langle e^{-2d/\lambda}\rangle/\lambda^2\eta_0 a$.
    }
    \label{Fig.14}
    \end{figure}

The vortex rf power $Q_{v}/a=B_0^2R_i/2\mu_0^2$ per unit surface area results in the additional surface
resistance $R_i$:
    \begin{equation}
    R_i=\frac{\phi_0^2\langle e^{-2d/\lambda}\rangle}{\lambda^2\eta_0 a}
    \Gamma_\omega(\sqrt{\Omega_p}),
    \label{ri}
    \end{equation}
where $a$ is a mean spacing between pinned vortices and
$\langle ... \rangle$ means averaging over the vortex positions $d$
in the direction perpendicular to the surface. Since
these positions must satisfy the stability condition $d>d_m$, the
main contribution to $R_i$ comes from vortices in the critical belt
$d_m<d<d_m+\lambda$ where $\langle e^{-2d/\lambda}\rangle$ is maximum.
For $\omega\tau_p\ll 1$, Eq. (\ref{ri}) simplifies to
    \begin{equation}
    R_i=\frac{\pi\mu_0^2\omega^2\ell^4\kappa^2\langle e^{-2d/\lambda}\rangle}{60a\rho_n\ln^2\tilde{\kappa}},
    \label{rilow}
    \end{equation}
where we used the Bardeen-Stephen $\eta_0$ and took $\epsilon
=\phi_0^2\ln\tilde{\kappa}/2\pi\mu_0\lambda^2$
for $\ell >\lambda$ or $\ln\kappa\to \ln (\ell/\xi)$ for $\xi\ll \ell \ll\lambda$ \cite{ehb}. These
results show that:

1. For $\omega\tau_p\ll 1$, the frequency dependence of $R_i$ is
similar to the BCS surface resistance, $R_{bcs}\propto
\omega^2\exp(-\Delta/T)$ at $T\ll T_c$. However, unlike
$R_{bcs}(T)$, the vortex contribution $R_i$ remains finite at $T\to
0$, so trapped vortices can contribute to the non-BCS excess surface
resistance, which has been often observed on many superconductors at
low temperatures. In this case even a few pinned vortices can result
in $R_i$ comparable to the exponentially small $R_{bcs}(T)$. This
scenario was first suggested by Rabinowitz \cite{rabin} who modeled
pinning by a phenomenological spring constant and did not considered
the critical depinning spacing $d_m$ due vortex attraction to the surface.
The account of a more realistic discrete pin structure in Fig. 13, and  the gradient
of the Lorentz force changes the frequency dependence of $R_i$ and results in new effects
discussed below.

2. $R_i$ increases significantly as the superconductor gets dirtier.
To evaluate this effect, we make a rough estimate $\langle
\exp(-2d/\lambda)\rangle\sim
\exp(-2d_m/\lambda)\lambda/a$, which takes into account the
main contribution to $R_i$ from vortices in the critical belt
$d_m<d<d_m+\lambda$. Taking $\exp(-2d_m/\lambda)$ from Eq.
(\ref{dm}) and using the asymptotic expansion
$K_1(z)=(\pi/2z)^{1/2}\exp(-z)$,  we obtain
    \begin{equation}
    R_i\sim \frac{\pi^2\mu_0^3\omega^2\ell^3\kappa^2\lambda^4f_p}{15a^2\phi_0^2\rho_n\ln^2\tilde{\kappa}}\sqrt{\frac{d_m}{\pi\lambda}}
    \label{rill}
    \end{equation}
Eq. (\ref{rill}) shows that $R_i\propto 1/a^2 $ is proportional to the trapped flux density, similar to the
Bardeen-Stephen flux flow resistivity.
In the dirty limit, $\lambda\simeq \lambda_0(\xi_0/\ell_i)^{1/2}$, $\xi\simeq (\xi_0\ell_i)^{1/2}$, and
$\rho_n\propto 1/\ell_i$, we obtain that the excess resistance $R_i\propto \rho_n^{3}$ increases rapidly as the resistivity
increases (here slowly varying logarithmic dependencies of $R_i$ on $\ell_i$ are neglected). This behavior of $R_i$ has
been observed on Nb cavities in which the change of $\rho_n$ at the surface was caused by a low-temperature
baking \cite{diss,gigi1,gigi2}.

3. There is a strong dependence of $R_i$ on the size $\ell$ of pinned vortex segments
because shorter segments have stiffer spring constants $\sim \epsilon/\ell^2$ and thus smaller
vibration amplitudes under rf field. If pinning centers are distributed randomly, $R_i(\ell)$ should be
therefore averaged over the distribution of the segment lengths:
    \begin{equation}
    \bar{R}_i(\omega)=\int_0^\infty R_i(\omega\eta_0\ell^2/2\epsilon)P(\ell)d\ell
    \label{avri}
    \end{equation}
where $R_i$ is given by Eq. (\ref{ri}), and $P(\ell)$ is a distribution
function of the segment lengths. For example, $P(\ell)$ can be taken in the form $P(\ell)=\ell_0^{-1}\exp(-\ell/\ell_0)$
used in the Granato-L\"{u}cke model of pinned dislocations, where $\ell_0$ is the mean segment length \cite{disl}.
In this case, Eqs. (\ref{ga}), (\ref{ri}) and (\ref{avri}) yield
    \begin{equation}
    \bar{R}_i(\Omega_p)=\frac{R_\infty}{\sqrt{\Omega_p}}\int_0^\infty e^{-z\Omega_p^{-1/2}} \Gamma_\omega(z)dz,
    \label{rif}
    \end{equation}
where $z=\ell/\ell_0$, $\Omega_p=\omega\eta_0\ell_0^2/2\epsilon$, and
$R_\infty=\phi_0^2\langle
e^{-2d/\lambda}\rangle/\lambda^2\eta a$ is the
high-frequency limit of $R_i(\omega)$. The resulting behavior of
$\bar{R}_i(\omega)$ is shown in Fig. 14: for low frequencies, the
contribution of weakly pinned large-length segments makes
$\bar{R}_i(\omega)$ higher than $R_i(\omega)$, while for high
frequencies, the contribution of strongly pinned small segments
makes $\bar{R}_i(\omega)$ lower than $R_i(\omega)$. The overall
behavior of $\bar{R}_i(\omega)$ resembles the power law dependence
$R_i\propto\omega^\beta$ with $\beta\simeq 0.5-0.7$, which has been
observed on Pb \cite{trap1} and Nb \cite{trap2} at 0.1-10 GHz.

To estimate the pinning time constant $\tau_p$, we first evaluate
the mean length of the vortex segment $\ell_0$. This can be done
from an estimate of the single-vortex pinning force balance $f_p\sim
J_c\phi_0\ell_0/\mu_0$, which express $\ell_0$ in terms of the
depinning critical current density $J_c$. For core pinning,
$f_p\simeq \zeta B_c^2\pi\xi^2/\mu_0$, where $\zeta$ accounts for
the change in the condensation energy by the pin due to variation of
$\delta T_c/T_c$ and the mean-free path \cite{s1}. Hence,
    \begin{equation}
    \ell_0\simeq \frac{\zeta\phi_0}{8\pi\mu_0\lambda^2J_c}
    \label{loo}
    \end{equation}
Taking $\lambda=40$nm and $J_c=10^9$A/m$^2$ for Nb at $T\ll T_c$
yields $\ell_0\sim 10^3\zeta\lambda$. The pinning relaxation time $\tau_p$
can then be obtained from Eqs. (\ref{taup}) and  (\ref{loo}):
    \begin{equation}
    \tau_p\simeq \frac{\zeta^2\phi_0^2}{128\pi^2\mu_0\rho_n\lambda^2\xi^2J_c^2\ln\tilde{\kappa}}
    \label{taupp}
    \end{equation}
Taking $\lambda\approx\xi\approx 40$nm, $J_c=10^9$A/m$^2$, and
$\rho_n=10^{-9} \Omega$m for Nb, we obtain $\tau_p [s]\sim
10^{-6}\zeta^2$. From the rf measurements of pinning
relaxation time $\sim 10^{-8}$ s in Nb \cite{trap2} we then
deduce $\zeta\sim 0.1$ and $\ell_0\sim10^2\lambda\simeq$ 4$\mu$m.
Here $\tau_p$ is rather sensitive
to the value of $\zeta$ determined by details of the order
parameter suppression at the pin.

It is instructive to compare $Q$ from an oscillating
vortex with $Q$ due to the rf electric field $E_i\simeq
B_0\omega\lambda\exp(-d/\lambda)$ induced in the fixed
normal vortex core. In the latter case the power $Q_v\simeq
\pi\xi^2E_i^2/\rho_n$ gives the surface resistance
$R_i\sim\pi\omega^2\lambda^4\exp(-2d/\lambda)/\mu_0^2\kappa^2a\rho_n$, which is by
the factor $\kappa^{-4}\ln^2\kappa\ll 1$ smaller than $R_i$ given by
Eq. (\ref{rilow}) for an oscillating vortex segment. Thus, for type-II superconductors
with $\kappa\gg 1$ considered in this paper, the inductive contribution \cite{rabin} is
negligible.

\subsection{Low-field nonlinear surface resistance and rf annealing of trapped magnetic flux}

So far we have considered $R_i$ independent of the rf field.
However, because $R_i$ is mostly determined by pinned vortices in
the critical belt $d_m<d<d_m+\lambda$, the excess resistance $R_i$
may become dependent on weak rf field due to an
increase of the critical distance $d_m$ as $B_0$ increases. This
effect is evident from Fig. 13, which shows that, because both the
image attraction force and the Meissner rf force increase as the
vortex moves closer to the surface during the negative rf cycle, the
vortex oscillations become asymmetric and shifted toward the
surface. Thus, some of pinned vortices at $d\simeq d_m$ can be
pushed out of the sample by rf field, which means that $d_m$
effectively increases as $B_0$ increases.

To calculate the effect of the rf field on the mean attraction force ${\bar f}(d)$ between a vortex and the surface,
we average Eq. (\ref{dynp}) in small vortex vibrations $\delta u(y,t)$:
    \begin{equation}
    {\bar f}=-\frac{\phi_0^2K_1(2d/\lambda)}{2\pi\mu_0\lambda^3}-
    \frac{f_\omega}{\lambda\ell}\int_{-\ell/2}^{\ell/2}\!dy\langle\sin\omega t\delta u\rangle_\omega .
    \label{ff}
    \end{equation}
In addition to the first static term in the r.h.s., $\bar{f}$ contains the rf contribution, in which $\langle ...\rangle_\omega$
 means time averaging over the rf period. Using $\langle\sin\omega t\delta u\rangle_\omega=
\mbox{Re}(\delta u_\omega)/2$ and Eq. (\ref{uom}) for the Fourier component $\delta u_\omega(y)$, we calculate
the integral in Eq. (\ref{ff}) as follows:
    \begin{equation}
    \mbox{Re}\!\int_{-\ell/2}^{\ell/2}\!\delta u_\omega\frac{dy}{2\ell}=-\frac{f_\omega\mbox{Re}}{\eta_0\omega\Omega_p^{1/2}}
    \left[\frac{\tan [(1-i)\Omega_p^{1/2}/2]}{1+i}\right]
    \label{integr}
    \end{equation}
Taking the real part of this expression, we reduce Eq. (\ref{ff}) to the
following equation for $d_m$:
    \begin{eqnarray}
    \frac{f_p}{\ell}=\frac{\phi_0^2 K_1(2d_m/\lambda)}{2\pi\mu_0\lambda^3}+
    \frac{\pi\ell^2B_0^2\tilde{\Gamma}_\omega[\Omega_p^{1/2}]}
    {2\mu_0\lambda\ln\tilde{\kappa}}e^{-2d_m/\lambda},
    \label{dmf} \\
    \tilde{\Gamma}_\omega(z)\frac{\sinh z-\sin z}
    {z^3(\cosh z+\cos z)}, \qquad\qquad
    \label{gammm}
    \end{eqnarray}
which defines the critical pinning depth $d_m$ as a
function of the rf amplitude and frequency. For
$\omega\tau_p\ll 1$ and $2d_m>\lambda$, we can use $\tilde{\Gamma}_\omega(0)\to 1/6$, and
$K_1(x)\simeq (\pi/2x)^{1/2}\exp(-x)$, in which case Eq. (\ref{dmf}) becomes
    \begin{eqnarray}
    \frac{f_p}{\ell}=\frac{\phi_0^2}{4\pi\mu_0\lambda^3}\sqrt{\frac{\pi\lambda}{d_m}}
    \left(1+\frac{B_0^2}{B_\phi^2}\right)e^{-2d_m/\lambda}
    \label{dlow} \\
    B_\phi=\frac{\phi_0}{\pi\lambda\ell}\left[\frac{\ln\tilde{\kappa}}{\tilde{2\Gamma}(\Omega_p^{1/2})}\right]^{1/2}
    \!\!\left(\frac{\pi\lambda}{d_m}\right)^{1/4}\quad
    \label{bf}
    \end{eqnarray}
Eq. (\ref{dlow}) differs from the static Eq. (\ref{dm}) by the factor
$(1+B_0^2/B_\phi^2)$, which becomes essential at rather low fields
$B_0\ll B_v$. Substituting $e^{-2d_m/\lambda}$ from Eq.
(\ref{dlow}) into Eq. (\ref{rilow}), we obtain the field dependence of
$R_i$:
    \begin{equation}
    R_i(B_0)=\frac{R_i(0)}{1+B_0^2/B_\phi^2},
    \label{rib}
    \end{equation}
where the zero-field $R_i(0)$ is given by Eq. (\ref{rilow}). The field
$B_\phi(\omega)=B_\phi(0)/\sqrt{6\tilde{\Gamma}_\omega}$ increases as $\omega$ increases,
approaching $B_\phi(\omega)=B_\phi(0)\Omega_p^{3/4}/\sqrt{6}$ for $\omega\tau_p\gg1$.

The decrease of $R_i(B_0)$ results from the field-induced shift of
the critical belt $d_m<d<d_m+\lambda$ away from the surface, where
the screening of the rf Meissner currents reduces vortex dissipation. As
a result, the rf field irreversibly pumps parallel vortices out of a
superconductor, resulting in a rf "annealing" of the field-cooled
trapped magnetic flux. The field dependence of $R_i(B_0)$ is
therefore hysteretic: if $B_0$ is first increased to a maximum value
$B_{0m}$ and then decreased back to zero, $R_i(B)$ on the ascending
branch decreases according to Eq. (\ref{rib}) and stay equal to
$R_i(B_{0m})$ on the descending branch as shown in Fig. 15. This
hysteretic behavior enables an experimental separation of the vortex
contribution to $R_i$ from reversible mechanisms due to
sound generation \cite{ris1}, dielectric losses in the substrate \cite{ris2,ris3} etc.

    \begin{figure}                  %FIG1
    \epsfxsize= 0.8\hsize
    \centerline{
    \vbox{
    \epsffile{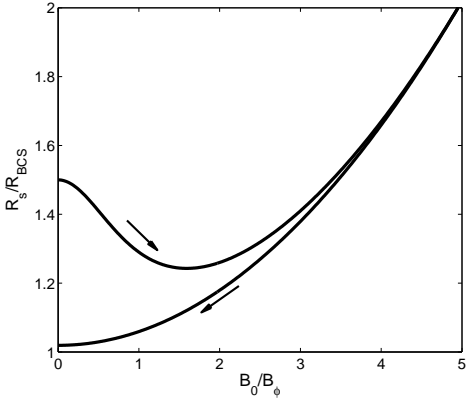}
    }}
    \caption{The hysteretic field dependence of $R_s(B_0)$ defined by Eq. (\ref{rsb}) for
    $\gamma=1$, $B_\phi=0.2B_c$, $R_i(0)=0.5R_{BCS}$. The descending branch is determined by
    Eq. (\ref{rsb}) in which $B_0$ in the last term stays at the maximum $B_0$ reached on the ascending branch.
    }
    \label{Fig.15}
    \end{figure}

The total surface resistance $R_s$ can be written as a sum of the
vortex component $R_i$ and the BCS resistance
    \begin{equation}
    R_s=R_{bcs}\left(1+\gamma\frac{B_0^2}{B_c^2}\right)+\frac{R_i(0)}{1+B_0^2/B_\phi^2},
    \label{rsb}
    \end{equation}
where we included the first nonlinear field correction to $R_{bcs}$
due to the Doppler shift of the quasiparticle spectrum, and also
pairbreaking and heating effects. In the clean limit the Doppler
contribution $\gamma_p \sim (\Delta/T)^2$ increases as $T$ decreases
\cite{nim1,nim2,nim3,nim4,ag}, while the heating component
$\gamma_h\propto R_{bcs}$ decreases as $T$ decreases \cite{ag}.
Because the vortex field $B_\phi$ is much smaller than $B_c$, the
total $R_s(B_0)$ can exhibit a nonmonotonic field dependence:
$R_s(B_0)$ first {\it decreases} as $B_0$ increases, reaching the
minimum $R_{sm}$ at $B_0=B_{min}$ and then increases at
$B_0>B_{min}$, as shown in Fig. 15. Here $B_{min}$ can be found from
Eq. (\ref{rsb}):
    \begin{equation}
    B_{min}^2=B_cB_\phi\sqrt{\frac{R_i(0)}{\gamma R_{bcs}}}-B_\phi^2
    \label{bmin}
    \end{equation}
The condition that the minimum in $R_s(B_0)$ occurs
is given by $B_{\min}^2>0$, that is,
    \begin{equation}
    \frac{R_i(0)}{\gamma R_{bcs}}>\frac{B_\phi^2}{B_c^2}
    \label{condi}
    \end{equation}
As illustrated by Fig. 15, the rf field cycling could reduce the
surface resistance by irreversibly pumping a fraction of trapped
flux out of the sample. The rf flux annealing considered in this
paper is somewhat analogous to a directional motion of magnetic flux
induced by transport ac current, resulting in a dc voltage on a
superconductor \cite{shake1,shake2,shake3}.

\section{Nonlinear hotspots in the surface resistance}

Localized dissipation due to vortex penetration or oscillation of
pinned vortices in thick films produces a long range temperature
distribution, which spreads out on the scale $\sim 2d/\pi$, much
greater than $\lambda$ (see Fig. 8). Even if these temperature
variations are weak, $\delta T({\bf r})=T({\bf r})-T_0 \ll
T_0$, they can nevertheless produce strong variations in the surface
resistance $R_s$ of the surrounding areas, resulting in nonlinear
contributions to $R_s$ with very different field and frequency
dependencies than $R_{bcs}(T,\omega)$. This effect comes from the
exponential temperature dependence of the BCS surface resistance,
    \begin{equation}
    R_{bcs}({\bf r})\propto \frac{\omega^2}{T}\exp\left [-\frac{\Delta}{T_0}+\frac{\delta T({\bf r})\Delta}{T_0^2}\right],
    \label{rbcs}
    \end{equation}
so that even weak variations $\delta T({\bf r})<T_0$ can produce
strong variations in $R_{bcs}({\bf r})$ at low temperatures,
$T_0^2<\delta T\Delta$. Indeed, substituting the surface temperature
distribution $\delta T(0,y)=(Q/\pi k)\ln\coth (\pi y/4d)$ from Eq.
(\ref{tx}) at $x=0$ into Eq. (\ref{rbcs}) we obtain:
    \begin{eqnarray}
    R_s(y)=R_{bcs}(T_0,\omega)\coth^\sigma\left(\frac{\pi y}{4d}\right),
    \label{hs} \\
    \sigma(B_0,T_0,\omega)=Q(B_0,T_0,\omega)\Delta(T_0)/\pi k(T_0) T_0^2
    \label{sigma}
    \end{eqnarray}
on the scales $|y|>r_0$ greater than the size $r_0$ of the heat
source. Here the exponent $\sigma$ is proportional to the dissipated
power $Q$, which depends on both $B_0$ and $\omega$.
For example, $Q\propto (B_0^2-B_v^2)^{\omega\tau_2}$ near the onset
of the single vortex penetration (see Eq. (\ref{q2}), or $Q\propto
B_0^2\Gamma_\omega(\Omega_p)$ for a pinned vortex near the surface
(see Eq. (\ref{qfin})). In turn, the dependencies of
$\sigma$ on $B_0$ and $\omega$ result in a nonlinear contribution to
the global surface resistance from sparse "hotspots" of size of the
film thickness around much smaller heat sources \cite{ag}. These hotspots contributions can have very
different dependencies on $B_0$ and $\omega$ as compared to the
field-independent $R_{bcs}\propto \omega^2$. For $\sigma<1$, the
total excess resistance $\delta \Re_s$ is insensitive to the power distribution $q(x)$, and can be obtained by
integrating Eq. (\ref{hs}) using a new variable $\varphi=\tanh^2(\pi y/4d)$:
    \begin{eqnarray}
    \delta\Re_s=\int_{-\infty}^{\infty}[R_s(y)-R_{bcs}]dy = \nonumber \\
    \frac{4d}{\pi}\left[\psi\left(\frac{1}{2}\right)-\psi\left(\frac{1-\sigma}{2}\right)\right]R_{bcs},
    \label{dr}
    \end{eqnarray}
where $\psi(x)=d\ln\Gamma(x)/dx$, and  $\Gamma(x)$ is the gamma-function \cite{abram}. For $\sigma\ll1$,
the expression in the brackets reduces to $\pi^2\sigma/4$, thus
    \begin{equation}
    \delta \Re_s=\pi d\sigma(B_0,\omega,T_0)R_{bcs},\qquad \sigma\ll 1.
    \label{dr1}
    \end{equation}
If $Q$ is due to pinned vortices, the correction $\delta \Re_s$ from
weak hotspots with $\sigma\ll 1$ is quadratic in $B_0$ and
proportional to $\omega^4$ for low frequencies $\omega\tau_p\ll 1$
[see Eq. (\ref{rilow})]. As $\sigma\to 1$, the function in the
brackets in Eq. (\ref{dr}) diverges logarithmically, indicating that
the spatial distribution of the power density $q(x)$, which cuts off
the logarithmic divergence in $T(x,y)$ in Eqs. (\ref{txy})
should be taken into account.

In the crossover region $\sigma\sim 1$, the behavior of $\delta
\Re(B_0,\omega)$ is sensitive to the details of the power density
distribution $q(x)$, but in the limiting case $\sigma\gg 1$ the main
contribution to $\delta \Re$ comes from the hottest region near the
heat source for which $\delta T(y)$ is given by Eq. (\ref{dty}).
Substituting Eq. (\ref{dty}) into
 Eq. (\ref{rbcs}), we obtain
    \begin{equation}
    \frac{\delta\Re_s}{R_{bcs}}\simeq \int_{-\infty}^{\infty}\!dy\left[\frac{16d^2}{\pi^2(y^2+r_0^2)}\right]^{\sigma/2}\!\!=
    r_0\left(\frac{4d}{\pi r_0}\right)^\sigma I_\sigma,
    \label{rsbig}
    \end{equation}
where $I_\sigma =
\sqrt{\pi}\Gamma[(\sigma-1)/2]/\Gamma(\sigma/2)\approx
(2\pi/\sigma)^{1/2}$ for $\sigma\gg 1$. The behavior of $\delta
\Re_s$ for $\sigma > 1$ changes radically as compared to $\sigma<1$:
instead of relatively weak power-law dependencies of $\delta \Re_s$
on $B_0$ and $\omega$ for $\sigma<1$, Eq. (\ref{rsbig}) predicts much
stronger exponential field and frequency dependencies of $\delta
\Re_s$ for $\sigma >$1, $r_0\ll d$.
The case of strong dissipation $\sigma>1$ can result from vortex
penetration amplified by grain boundaries \cite{gb1,gb2}, surface
topography \cite{hasan}, local enhancements of $R_{bcs}$ due to
impurity segregation etc.

The mean surface resistance ${\bar R}_s$ averaged over all hotspot contributions is given by
    \begin{equation}
    {\bar R}_s=R_{bcs}+{\bar R}_i+\sum_n \delta \Re_s ({\bf r_n})/A.
    \label{globr}
    \end{equation}
Here the averaged residual resistance $\bar{R}_i$ results from
either pinned vortices or other mechanisms, the last term in the
r.h.s. is due to the effect of vortex dissipation on the BCS
resistance, ${\bf r_n}$ are the coordinates of the sparse (thermally
noninteracting) hotspots, and $A$ is the surface area exposed to the
rf field.  As shown above, $\delta\Re_s$ can have very different
temperature and frequency dependencies as compared to $R_{bcs}$, so
the hotspot contribution can strongly affect the dependencies of
global surface resistance ${\bar R}_s$ of $\omega$ and T,
particularly at low temperatures, where  $R_{bcs}$ is exponentially
small. Moreover, the last 2 terms in the r.h.s. of Eq. (\ref{globr})
can bring about a strong dependence of ${\bar R}_s$ on the rf amplitude,
which can be well below the field $\sim B_cT/T_c$
of intrinsic nonlinearities of the BCS surface resistance due to the
Doppler shift of quasiparticle energies \cite{ag,nim1,nim2,nim3,nim4}.

\subsection{Thermal instabilities ignited by hotspots}

In the previous section we considered the rf power as a function on the
bath temperature $T_0$. However, $Q$ is actually determined by the local temperature $T_m$, which
should be calculated self-consistently from the heat balance condition.
We consider the case, for which the mean spacing between
hotspots $L_i$ is shorter than the thermal length $L_\theta =
(dk/\alpha_\theta)^{1/2}$, over which $\delta T(\bf{r})$ decays away from a single hotspot.
Here $\alpha_\theta = k\alpha_K/(d\alpha_K
+k)$ is the effective thermal resistance across the film which accounts
for the resistance $d/k$ due to thermal conductivity plus the
interface thermal resistance $1/\alpha_K$, where $\alpha_K$ is the
Kapitza heat transfer coefficient. For $L_i\ll L_\theta$, thermal
fields of hotspots overlap and the temperature $T_m$ along
the surface becomes uniform. In this case the thermal balance
equation takes the form
    \begin{equation}
    (T_m-T_0)\alpha_\theta=\frac{B_0^2}{2\mu_0^2}[R_i+R_{bcs}(T_m)],
    \label{qsam}
    \end{equation}
which determines self-consistently both the rf dissipated power and
the maximum temperature $T_m$ as functions of $B_0$ and $\omega$. It
is convenient to express $B_0(T_m)$ as a function of $T_m$ from Eq.
(\ref{qsam}):
    \begin{equation}
    B_0^2=\frac{2\mu_0^2(T_m-T_0)\alpha_\theta}{R_i+R_0\exp[(T_m-T_0)\Delta/T_0^2]}
    \label{boo}
    \end{equation}
Here we took into account the most essential exponential temperature
dependence of the BCS surface resistance, where $R_0=R_{bcs}(T_0)$,
and the residual resistance $R_i$ due to trapped vortices is assumed
temperature-independent for $T\ll T_c$. The function $B_0(T_m)$ has
a maximum at:
    \begin{equation}
    T_m-T_0=\frac{T_0^2}{\Delta}+\frac{B_0^2R_i}{2\mu_0^2\alpha_\theta},
    \label{tmaxx}
    \end{equation}
giving the critical overheating  $T_m-T_0\ll T_0$ above which a thermal instability develops.
From Eqs. (\ref{tmaxx}) and (\ref{boo}),
we obtain the equation for the maximum field $B_p$:
    \begin{equation}
    \frac{R_0B_p^2e\Delta}{2\mu_0^2T^2\alpha_\theta}\exp\left(\frac{R_iB_p^2\Delta}{2\mu_0^2T^2\alpha_\theta}\right)=1
    \label{eqbp}
    \end{equation}
The thermal balance Eq. (\ref{qsam}) has
solutions $T_m(B_0)$ only if the rf amplitude is below the breakdown
field $B_p$. For $B_0>B_p$, the thermal runaway occurs because the
heat generation grows faster than the heat flux to the coolant as
$T_m$ increases. This situation is analogous to
combustion in chemical systems \cite{chem} or thermal quench in semiconductors,
normal metals or superconductors \cite{gm}.

    \begin{figure}                  %FIG1
    \epsfxsize= 0.8\hsize
    \centerline{
    \vbox{
    \epsffile{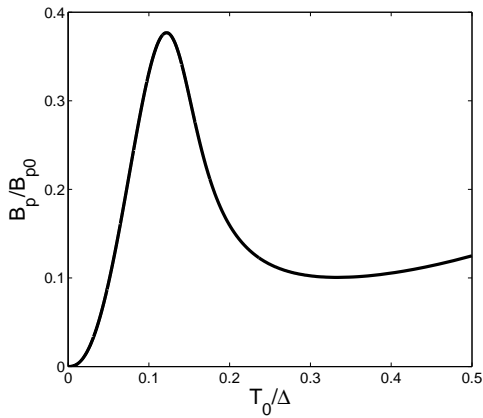}
    }}
    \caption{The temperature dependence of $B_p(T_0)$ calculated from Eq. (\ref{eqbp})
    for $R_{bcs}(T_0)=R_{n0}(\Delta/T_0)\exp(-\Delta/T_0)$, $R_i/eR_{n0}=2\times 10^{-5}$, $\alpha_\theta
    = \alpha^\prime T_0^s$, $s=3$, and $B_{p0}^2=2\mu_0^2\alpha^\prime T_0^{s+1}/R_{n0}$.
    }
    \label{Fig.16}
    \end{figure}

For $R_i\ll R_0$, the exponential term in Eq. (\ref{eqbp}) can be neglected and the breakdown field is given by\cite{ag}
    \begin{equation}
    B_p=\mu_0\left(\frac{2\alpha_\theta
    T_0^2}{R_0e\Delta}\right)^{1/2}
    \label{bpo}
    \end{equation}
The temperature dependence of $B_p(T_0)$ can be obtained taking
$R_0(T_0)=R_{n0}(\Delta/T_0)\exp(-\Delta/T_0)$, $\alpha_\theta  =
\alpha^\prime T_0^s$, whence $B_p(T_0)\propto
T_0^{(3+s)/2}\exp(-\Delta/2T_0)$ reaches a minimum at
$T_{min}=\Delta/(s+3)$ and increases as $T_0$ decreases below
$T_{min}$. However, for lower temperatures, $R_{bcs}(T_0)$ becomes
smaller than $R_i$ in which case the exponential term in Eq.
(\ref{eqbp}) dominates, and $B_p(T_0)\sim \mu_0(2\alpha_\theta
T_0^2/R_i\Delta)^{1/2}\propto T_0^{1+s/2}$ decreases as $T_0$
decreases. The behavior of $B_p(T_0)$ is shown in Fig. 16.  The maximum in $B_p(T)$ at
the optimum temperature $T_{max}$ separates the regimes controlled
by the BCS surface resistance at $T_0>T_{max}$ and by hotspots due
to frozen flux or other mechanisms of residual resistance at
$T_0<T_{min}$. For Nb, ($\Delta\approx 18$K), the optimum
temperature in Fig. 16 corresponds to $T_{max}\simeq 2$K, while for
Nb$_3$Sn ($\Delta\approx 36$K), $T_{max}\simeq $4K, if $R_i/R_{n0}$
is the same for both materials. The above consideration based on the
linear BCS surface resistance assumes that the breakdown field $B_p$
is much smaller than the field $\simeq TB_c/T_c$ at which the
intrinsic nonlinearities in $R_s$ become important. A significant
increase of the isothermal $R_s(B_0)$ due to these nonlinearities
can strongly affect the thermal breakdown \cite{ag}, limiting $B_p$
by the thermodynamic critical field $B_c$.

\section{Discussion}

The results presented above show that the breakdown of the Meissner state by strong rf fields
involves supersonic vortex penetration through the surface barrier weakened by defects, the jumpwise
LO-type instability and high dissipation even for single vortices. Such dissipation results in thermal
retardation effects and hotspots igniting the explosive thermal instability due to the exponential temperature
dependence of the surface resistance. These effects are precursors for the avalanche vortex penetration and
dendritic thermo-magnetic instabilities \cite{den1,den2}.

At the onset of vortex penetration pinning forces are much weaker than the driving forces
of the rf Meissner currents. Yet, as the vortex moves away from the surface by the distance $\sim\lambda$,
the Lorentz force decreases exponentially, so the vortex can be trapped by
pinning centers. Such vortices trapped in the thin surface layer of rf field penetration
during breaking through the surface barrier or field cooling of the sample can result in a temperature-independent
residual surface resistance. However, because pinning centers are distributed randomly, the rf power dissipated by
pinned vortices $Q(u)$ varies very strongly because of the exponential sensitivity of
$Q\propto\exp(-2u/\lambda)$ to the vortex position.  This effect results in
hotspots of vortex dissipation, which peaks for vortices spaced from the surface by distances close to the minimum distance
$d_m$, for which pinning forces can prevent vortex annihilation at the surface at low fields $B_0\ll B_v$.
The field dependence of $d_m$ causes rf flux annealing in which vortices are irreversibly pushed out from the surface
layer. This effect results in a nonlinear hysteretic dependence of $R_i(B)$ at low fields, $B_0\ll B_c$,
which may pertain to the puzzling decrease of the surface resistance at low fields $B_0\sim 3-20$mT, which has been often observed
on Nb \cite{hasan,gigi1} and other superconductors \cite{ris2,ris3}. Yet, Eq. (\ref{rib}) describes
well the field dependence $R_i\propto 1/B_0^2$ observed on Nb cavities \cite{gigi1}.

Besides the field and frequency dependencies of $R_i(B)$, another manifestation of the vortex pinning
mechanism is the hysteretic behavior shown in Fig. 14 due to the rf annealing of the trapped flux. However,
this mechanism caused by the gradient of the Lorentz force is only effective for the vortex segments parallel
to the surface and does not affect vortex segments perpendicular to the surface.  The segments of pinned vortices
perpendicular to the surface generally give a field-independent contribution to $R_i$, however if these segments belong to
the vortex semi-loop trapped at the surface, the rf Lorentz force gradient acting on the parallel component of the loop can
eventually push the whole loop toward the surface where it shrinks and annihilates. In this case the rf annealing decreases
$R_s$, eliminating some of the hotspots caused
by trapped vortices. This is illustrated by Fig. 17, which shows 3 different type of vortices trapped at the surface.
Vortex 1 cannot be pushed out by the rf field because only a small segment of it $\sim\lambda$ is exposed to the rf Lorentz force,
while the rest part is pinned in the bulk. Vortex semi-loop 2 can be pushed out by the rf field, as
discussed above. Vortex 3 has a parallel segment, which however cannot annihilate at the surface because it
is held back by other pinned segments which are beyond the surface layer of the rf field penetration.

 \begin{figure}                  %FIG1
    \epsfxsize= 0.3\hsize
    \centerline{
    \vbox{
    \epsffile{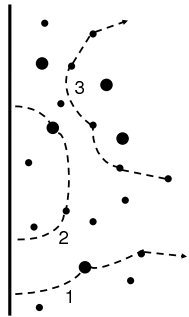}
    }}
    \caption{Vortices (shown as dashed lines) trapped near the surface by pinning centers (black dots).
    }
    \label{Fig 17}
    \end{figure}

Penetration and trapping of even single vortices at low temperatures
can significantly increase the exponentially small $R_s$, which in
turn, decreases the thermal breakdown field $B_p$. For example, flux
trapped during field cooling in the Earth magnetic field $B_i\simeq
40 \mu$T, corresponds to the mean intervortex spacing
$a=(\phi_0/B_i)^{1/2}\simeq 7\mu$m. To estimate $R_i$ for such
trapped flux, we use Eq. (\ref{ri}) for strong pinning at
$\omega\tau_p<1$, taking $\langle \exp(-2d/\lambda)\rangle\sim
\lambda/a$ and $d_m\sim \lambda$. Hence,
    \begin{equation}
    R_i\simeq \frac{\omega^2\tau_p^2}{30}R_i(\infty), \qquad R_i(\infty)=\frac{\rho_nB_i}{\lambda B_{c2}},
    \label{rist}
    \end{equation}
where $R_i(\infty)$ is the high frequency limit of $R_i(\omega)$.
For Nb, taking $B_{c2}=400$ mT, $\lambda=40$ nm,
$\rho_n=10^{-9}\Omega$m, and $B_i=40 \mu$T, we obtain
$R_i(\infty)=2.5 \mu\Omega$, much higher than the typical values
$R_i\sim 10 n\Omega$ observed on high-purity Nb \cite{gigi1}. For
Nb$_3$Sn, with $\rho_n=0.2\mu\Omega$m, $\lambda\simeq 90$ nm, we
obtain $R_i(\infty)\simeq 3.9 \mu\Omega$. As
follows from Eq. (\ref{rist}), pinning reduces $R_i$ by the factor
$(\omega\tau_p)^2/30$, so $R_i\simeq 10 n\Omega$ corresponds to
$(\omega\tau_p)^2\sim 0.1$, or $\tau_p\simeq 0.5$ ns for
$\omega/2\pi=1$ GHz.  By contrast, the pinning time constant
$\tau_p\sim 10^{-8}$ s  measured by Pioszyk et al. \cite{trap2}
seems to indicate that their Nb sample was in the weak pinning
limit, $(\omega\tau_p)^2\gg 1$, for which $R_i\simeq R_i(\infty)$ is
in agreement with the measured $R_i(0.5 Oe)\simeq 2 \mu\Omega$
and the estimate from Eq. (\ref{rist}).

Introducing dense
pinning structures in the surface layer of the rf field penetration can
therefore impede vortex oscillations and significantly reduce the part of
$R_i$ caused by trapped flux, particularly for vortex loops like 1
and 3 in Fig. 17 which are not affected by the rf flux annealing.  Because
pinning is only effective if $\omega\tau_p<1$, decreasing $\tau_p$
in Eq. (\ref{taup}) implies reducing the pin spacing. At the same time, a more effective rf flux
annealing requires both weak and dense pins (small $f_p$ and $\ell$ in Eq. (\ref{rill})). Overall,
$R_i$ can be reduced by decreasing the relaxation time constant $\tau_p$,
which can be done not only by decreasing the pin spacing $\ell$ but also by optimizing the
mean-free path at the surface. Indeed, Eq.
(\ref{taup}) shows that $\tau_p$ is larger for higher-$\kappa$
superconductors because of the softening of the vortex line tension
$\epsilon$, although this effect can be offset by a higher normal
resistivity. For example, the ratio $\tau_{p1}/\tau_{p2}=
\kappa_1^2\rho_{n2}/\kappa_2^2\rho_{n1}\propto \rho_{n1}/\rho_{n2}$ for two different materials 1 and 2
(or two different mean free paths $\ell_i$) but
the same $\ell$, shows that pinning becomes less effective for a dirtier surface.
Furthermore, comparing Nb with $\kappa_1=1$, $\rho_{n1}=10^{-9} \Omega$m and Nb$_3$Sn
 with $\kappa_2=30$ and $\rho_{n1}=0.2 \mu\Omega$m, we obtain
 $\tau_p^{Nb}/\tau_p^{Nb_3Sn}\sim 1/5$. Thus, reduction of $R_i$ by pinning turns out to be
 somewhat more effective in Nb, although this conclusion can be strongly affected by impurities, as discussed above.

The results of this work show that reducing vortex dissipation is an
important problem in achieving ultimate pairbreaking breakdown
fields in superconductors. In particular, a significant progress has
been made in increasing $R_s$ and $B_p$ by low-temperature annealing
of Nb cavities which enables tuning the impurity concentration,
nanoscale oxide layers and hotspot distribution on Nb surface
\cite{gigi1,gigi2,gigi3}. Another possibility in raising the
ultimate breakdown fields is to use thin film
superconductor-insulator-superconducting (SIS) multilayer coating
with high-$B_c$ films of thickness $d<\lambda$ to significantly
increase $B_{c1}$ and delay the field onset of vortex penetration
\cite{ag1}. Moreover, the SIS coating may suppress the LO
instability by decreasing the vortex flight time through the film
and providing strong pinning due to magnetic interaction of the
vortex with the film surfaces. The SIS multilayer coating of Nb
cavities may enable increasing the ultimate breakdown fields $B_p$
above $B_c^{Nb}$ by taking advantage of A15 superconductors
\cite{a15} or MgB$_2$ \cite{mgb2,mgb2r} with $B_c>B_c^{Nb}$ and
potentially lower $R_{bcs}$.

\section{Acknowlegements}

The work at NHMFL is supported by NSF Grant DMR-0084173 with support
from the state of Florida. The work at the Jefferson Laboratory is supported by the Jefferson Science Associates, LLC under
DOE Contract No. DE-AC05-060R23177.

\appendix

\section{Temperature of a moving vortex}

Eq. (\ref{tdeq}) can be written in the dimensionless form:
    \begin{equation}
    \dot{\theta}=\nabla^2\theta - \theta+\beta(\theta_m)s^2(t)f[x-u(t),y],
    \label{a1}
    \end{equation}
where $\theta=(T-T_0)/(T_c-T_0)$, and
the time, coordinates and vortex velocity are normalized by the
thermal scales $t_\theta = C/\alpha$, $L_\theta =
(k/\alpha)^{1/2}$ and $v_\theta = L_\theta/t_\theta$,
respectively,
$\beta(\theta_m)=\eta_0(\theta_m)v_\theta^2/(T_c-T_0)k$, and
$s(t)=v/v_\theta$ is the dimensionless vortex velocity. The Fourier
transform of Eq. (\ref{a1}) results in the following equation for
the Fourier components, $\theta_p(t)=\int \theta(x,y,t)\exp(-i{\bf
pr})d^2{\bf r}$:
    \begin{equation}
    \dot{\theta_{\bf p}}+(1+p^2)\theta_{\bf p}=f_{\bf p}g(t)e^{-ip_xu(t)}
    \label{a2}
    \end{equation}
where $g(t)=\beta[\theta_m(t)]s^2(t)$. The solution of Eq.
(\ref{a3}) is:
    \begin{equation}
    \theta_{\bf p}(t)=f_{\bf p}\int_0^\infty e^{-(1+p^2)t'-ip_xu(t-t')}g(t-t')dt'
    \label{a3}
    \end{equation}
For steady-state vortex oscillations, the rf
field was turned on at $t_i=-\infty$, and $v(t)$ in Eq. (\ref{a3})
accounts for all oscillations preceding the time $t$. For
$f(r)=\pi^{-1}\xi_1^{-2}\exp(-r^2/\xi_1^2)$, and $f_{\bf p}=\exp
(-p^2\xi_1^2/4)$, the inverse Fourier transform of Eq. (\ref{a3})
gives
    \begin{equation}
    \theta({\bf r},t)=\frac{1}{\pi}\int_0^\infty\frac{dt'g(t-t')}{4t'+{\tilde \xi}^2}
    e^{-t'-\frac{[x-u(t-t')]^2+y^2}{{\tilde \xi}^2+4t'}},
    \label{a4}
    \end{equation}
Eq. (\ref{a4}) was obtained for an infinite sample. For a thermally-insulated or ideally cooled surface at $x=0$, Eq.
(\ref{a4})  can be modified using the method of images:
    \begin{eqnarray}
    \theta({\bf r},t)=\frac{1}{\pi}\int_0^\infty\frac{dt'g(t-t')}{4t'+\tilde{\xi}^2}\times \nonumber \\
    e^{-t'-\frac{y^2}{\tilde{\xi}^2+4t'}} \bigl[e^{-\frac{[x-u(t-t')]^2}{\tilde{\xi}^2+4t'}}
    \pm e^{-\frac{[x+u(t-t')]^2}{{\tilde\xi}^2+4t'}}\bigr],
    \label{a5}
    \end{eqnarray}
where the plus and minus signs in the parenthesis correspond to the Dirichlet $[\partial_x\theta(0,y,t)=0]$ and the Neumann
$[\theta(0,x,t)=0]$ boundary conditions at the surface, respectively.
The core size is typically much smaller than $L_\theta$, so the dynamic equation for the core
temperature $\theta_m(t)$ can be obtained from the self-consistency condition $\theta_m(t)=\theta(u(t),0,t)$,
resulting in Eq. (\ref{tm}).

Next we consider the steady-state temperature distribution $T(x,y)$ averaged over high-frequency vortex
oscillations, at $2\pi\omega t_\theta >> 1$. In this case
$T(x,y)$ satisfies the static thermal diffusion equation
    \begin{equation}
    k\nabla^2 T+q(x)\delta(y)=0
    \label{a6}
    \end{equation}
where $q(x)$ is the mean power density distribution along the vortex
trajectory. We solve Eq. (\ref{a6}) for a film of thickness $d$ with
the boundary conditions: $\partial_x T(x,y)=0$ on the
thermally-insulated surface $x=0$ where vortex dissipation is localized,
and $T(x,y)=T_0$ at the opposite surface, $x=d$ kept at $T_0$. By symmetry, this geometry has the
same $T({\bf r})$ as in a film of thickness $2d$ with isothermal
boundary conditions $T(\pm d,y)=T_0$, and the heat source in
the middle at $x\approx 0$. In this case Eq. (\ref{a6}) can be
solved using the Green function
    \begin{equation}
    G({\bf r},{\bf r'})=\frac{1}{4\pi k}\ln\frac{\cosh\frac{\pi(y-y')}{2d}+
    \cos\frac{\pi(x+x')}{2d}}{\cosh\frac{\pi(y-y')}{2d}-\cos\frac{\pi(x-x')}{2d}},
    \label{a7}
    \end{equation}
which gives the distribution of $\delta T({\bf r})=T({\bf r})-T_0$:
    \begin{equation}
    \delta T({\bf r})=\frac{1}{2\pi k}\int_0^dq(x')\ln\frac{\cosh\frac{\pi y}{2d}+
    \cos\frac{\pi(x+x')}{2d}}{\cosh\frac{\pi y}{2d}-\cos\frac{\pi(x-x')}{2d}}dx'.
    \label{a8}
    \end{equation}
If the length of dissipation source $\sim r_0$ is much smaller than the film thickness,
$\delta T({\bf r})$ around the source at $(x^2, y^2) \ll d^2$ reduces to
    \begin{equation}
    \delta T({\bf r})=\frac{1}{2\pi k}\int_0^\infty q(x')\ln\frac{16 d^2}{\pi^2[y^2+(x-x')^2]}dx'.
    \label{a9}
    \end{equation}
Next, we take a rectangular approximation  $q(x) =
q_0$ for $x<r_0$ and $q(x)=0$ for $x>r_0$, where $r_0$ is a characteristic size of the dissipation source, so that
$q_0r_0 = Q$ gives the total power Q. In this case the distribution of $\delta T(0,y)$
along the surface becomes
    \begin{equation}
    \delta T(y)=\frac{q_0}{2\pi k}\int_0^{r_0}\ln\frac{16 d^2}{\pi^2(y^2+x'^2)}dx',
    \label{a10}
    \end{equation}
which yields after integration:
    \begin{equation}
   \delta T(y)=\frac{Q}{2\pi k}\left[\ln\frac{16d^2}{\pi^2(y^2+r_0^2)}+2-\frac{2y}{r_0}\tan^{-1}\!\frac{r_0}{y}\right]
    \label{a11}
    \end{equation}
The maximum $\delta T_m=\delta T(0,0)$ is given by:
    \begin{equation}
    \delta T_m =\frac{Q}{\pi k}\left(\ln\frac{4d}{\pi r_0}+1\right)
    \label{a12}
    \end{equation}

\section{Multivalued friction force}

Eq. (\ref{dyni}) describes an overdamped vortex driven
by the force $F(u,t)$ balanced by a nonlinear friction force $\eta
v/(1+v^2/v_0^2)$. The force balance equation has either two or no
roots, as shown in Fig. 17. The velocity $v_-(F)$ for the left
intersection point vanishes at $F=0$ and increases up to $v_0$ as $F$
increases. The velocity for the right intersection point $v_+(F)$
decreases as F increases.
    \begin{figure}                  %FIG1
    \epsfxsize= 0.75\hsize
    \centerline{
    \vbox{
    \epsffile{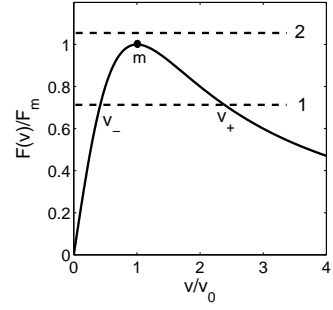}
    }}
    \caption{ Graphic solution of the equation $F(v)=F$, where
    $F(v)=\eta_0v/(1+v^2/v_0^2)$. The branches $v_\pm$ correspond to
    the $\pm$ signs in Eq. (\ref{qeq}), and the dashed lines show
    the Lorentz force $F$ for particular values of $u$ in the
    stable region (1) and the unstable region (2).
    }
    \label{Fig.18}
    \end{figure}
If $F<F_m$ the branch $v_-(F)$ describes all smooth parts of the
vortex trajectory and also provides $v_-(0)=0$ for the initial
condition $u(0) = 0$. For $F > F_m$, the vortex jumps from $x=u_1$
to the point $x=u_2$ where friction is able to balance the drive.
Here $u_{1,2}$ are defined by the condition
$F(u_{1,2},t_{1,2})=F_m$. The branch $v_-(F)$ describes all smooth
parts of $u(t)$ provided that $\dot{F}(u_{1,2},t) < 0$, so that
$F(t)$ always decreases below $F_m$ after the jump.

For high rf frequencies or strong vortex core overheating, there are
situations, for which $\dot{F} > 0$ after the jump. For example, if
$u > \lambda$, the term with the $K(x)$ in F can be neglected, and
$\dot{F} = v_0\partial_xF +\partial_tF$ becomes:
    \begin{equation}
    \dot{F}=(-v_0/\lambda+\omega\cot\omega t)F,
    \label{b1}
    \end{equation}
which tends to become positive at higher frequencies. In this case
the vortex cannot jump to the point where $F(u,t)=F_m$, since the
friction force cannot balance $F$ if $\dot{F} >0$, because
$F(u,t)$ keeps increasing above $F_m$ so $v(t)$
becomes greater than $v_0$. Thus, for $\dot{F}
> 0$, friction can only stop the vortex jump if $v
> v_0$, thus we have to consider the branch $v_+(F)$ as well.
We should therefore construct the trajectory, for which the vortex
first jumps at $t = t_i$ to the new point $x = u_i$ and acquires the
velocity $v_i>v_0$. After that $v(t)$ continuously decreases from
$v_i> v_0$ at $t = t_i$ to $v_0$ at $t = t_0$, as described by the
first order differential equation $\dot{u}=v_+[F(u,t)]$. Here $u_0$
and $t_0$ are determined by the equations:
    \begin{equation}
    F[u_0,t_0]=F_m,\qquad v_0\partial_uF+\partial_tF=0,
    \label{b2}
    \end{equation}
which state that the $\dot{F}$ should change sign as the vortex
reaches the maximum friction force at the critical velocity $v-0$.
Indeed, if $\dot{F}$ changes sign at any point of the descending
branch $v_+(F)$, the vortex cannot not reach $v_0$, so $v(t)$ passes
through a minimum at some $v>v_0$ and then starts accelerating
continuously. On the other hand if the vortex reaches $v_0$ at
$\dot{F}> 0$, the jump instability occurs. Therefore Eq. (\ref{b2})
provides the only way for a stable switch from the descending branch
of $v_+(F)$ for $t_i < t < t_0$, to the ascending branch of
$v_-(F)$. Once $t_0$ and $u_0$ are found from Eq. (\ref{b2}), the
coordinate of the jump, $u_i$, can be calculated for the vortex
going backward in time, taking the initial condition $u = u_0$, and
$F = F_m$ and then, solving the equation ${\dot
u}=v_+[F(t_0-\tilde{t})]$ with the "negative time"
$\tilde{t}=t_0-t_i$, from $\tilde{t}=0$ to $\tilde{t}= t_0 - t_i$.

\end{document}